# The Propulsion Performance of Dual Heaving Foils in Tandem, with Spanwise Flexibility

Wendi Liu[1,*], Alex Skillen[2], Wei Wang[1], Charles Moulinec[1] and David R. Emerson[1]

1. Scientific Computing Department, Science and Technology Facilities Council, Daresbury Laboratory, Warrington WA4 4AD, UK;
2. Department of Fluids and Environment, School of Engineering, The University of Manchester, UK.

**ABSTRACT**

The impact of spanwise flexibility on the propulsion performance of two foils arranged in tandem and subjected to a prescribed sinusoidal heaving motion has been studied at a Reynolds number of 100. This comprises a wide range of natural frequencies of the flexible foil, covering pre-resonance, resonance, and post-resonance regimes, with respect to the heaving frequency. The propulsion performance of the foils under different Strouhal numbers, with either fixed reduced frequency or fixed natural frequency, has also been investigated. We observed that the propulsion efficiency, thrust, and power consumption drop, but the tip deformation coefficient increases to its maximum when operating at resonance. The propulsion efficiency of the flexible dual foils increases significantly in the pre-resonance regime at small Strouhal numbers. Both thrust force and power consumption increase with the reduced frequency during the pre- and post-resonance regimes. An instability phenomenon is observed in the instantaneous tip deformation coefficient of the flexible dual foils in the post-resonance regimes. We also find that the dual heaving foils, with a larger resonance Strouhal number, achieve a greater thrust increment compared with that of rigid dual heaving foils when working at the pre-resonance regimes. Meanwhile, dual heaving foils with a smaller resonance Strouhal number can achieve a larger propulsion efficiency increment compared with that of rigid dual heaving foils when working at small Strouhal numbers.

## 1. Introduction

Many animals use flapping wings or fins for their propulsion and manoeuvre. To improve understanding of this propulsive motion, the *heaving foil has been* widely adopted model as a simplified representation of flapping. In the early 20th century, Knoller (1909) and Betz (1912) independently observed there was an effective angle of attack created by a flapping wing in generating lift and thrust forces, which is known as the Knoller–Betz effect. Since then, many studies have been carried out that reveal the mechanism of heaving foil propulsion, both experimentally and numerically (Katzmayr 1922, Garrick 1936, Jones *et al.,* 1998, Floryan *et al.,* 2017).

When two foils are involved, common configurations include parallel or tandem placements. A typical scenario where this happens is when two fins or wings are close to each other e.g. when fish swim in schools or birds fly in groups. Both parallel and tandem placements are possible in this scenario. For parallel configurations, the chord lines of the two foils are parallel to each other. Increased fluid forces will occur for the parallel configured heaving foils when the gap between the two foils becomes small (i.e. the two foils more closer to each other) during the heaving cycle (Koutsogiannakis *et al.,* 2019). Tandem configurations have one foil located downstream of the other foil. The wake of the upstream foil can have a strong impact on the downstream foil. Tandem configured wings generate distinct flow features due to forewing and hindwing interactions (Shyy *et al.,* 2010, Azuma 2012).

---

[*] Corresponding author. E-mail address: wendi.liu@stfc.ac.uk

The flexibility of a foil can be broadly classified into two categories, namely chordwise and spanwise flexibility. Previous studies that have focused on chordwise flexibility include (Heathcote *et al.,* 2004, Pederzani and Haj-Hariri 2006, Heathcote and Gursul 2007, Zhu 2007, Gulcat 2009). It can be concluded from these studies that chordwise flexibility can affect the force generation in three main ways: 1) changing the foil's effective angle of attack, 2) changing the net force direction and 3) changing the effective geometry of the foil (Shyy *et al.,* 2010). There are comparatively fewer studies on spanwise flexibility. Liu and Bose (1997) studied the effects of spanwise flexibility on the propulsive performance of a 3-D foil under prescribed heaving and pitching motion using the panel method. They found that the phase difference between the spanwise elastic motion and the prescribed motion have a significant impact on the thrust generation and the propulsion efficiency. Zhu (2007) employed a fully coupled fluid-structure interaction (FSI) model, which used the boundary-element method for the fluid domain and a 2-D nonlinear thin-plate model for the structural domain, to investigate the effect of structural flexibility on the performance of a 3-D rectangular thin foil under prescribed flapping (i.e. combined pitching and heaving) motion, with both chordwise flexibility and spanwise flexibility. They examined the thin foil in both air and water and compared the outcome. They found the foil's deformation was similar to the no fluid case when the foil was in the air, which is defined as inertia-driven deformation. When the foil was in water, there was significant interaction between the water and the foil, resulting in fluid-driven deformation. They found that chordwise flexibility with fluid-driven deformation increases efficiency by reducing the effective angle of attack, while spanwise flexibility with inertia-driven deformation increases thrust without sacrificing efficiency. However, their conclusions are limited to large Reynolds numbers.

One of the most influential studies on the spanwise flexibility of a foil propulsor was done by Heathcote *et al.,* (2008). They performed a series of water tunnel experiments to investigate the impact of spanwise flexibility on the thrust, lift and propulsion efficiency of a rectangular foil with pure heaving motion. Foils with three levels of flexibility were constructed, which are the "inflexible" foil constructed from nylon and stiffened with two 8 mm diameter steel rods, the "flexible" foil constructed from polydimethylsiloxane (PDMS) rubber and stiffened with 1 mm stainless steel sheet and the "highly flexible" foil constructed from PDMS and stiffened with aluminium sheet. A prescribed sinusoidal heaving motion was imposed at the leading edge of the foil root. The integrated foil thrust coefficient and tip displacement were measured. Their results show that the "flexible" foil yields a small increase in thrust coefficient and a small decrease in power consumption, resulting in higher propulsion efficiency compared with that of the "inflexible" foil. The trailing-edge vortices of the "flexible" foil are also found to be stronger than that of the "inflexible" foil. However, for the "highly flexible" foil, the thrust coefficient and the propulsion efficiency were significantly reduced compared with that of the "inflexible" foil. A large phase difference in the tip displacement between the two tips of the foil was observed, resulting in the root and tip moving in opposite directions for a significant portion of the heaving cycle. A weak and fragmented vorticity pattern was observed for the trailing-edge vortices of the "highly flexible" foil, in which vortices with opposing rotating directions were shed from the foil root and the tip, respectively. Numerical simulations of these experiments were carried out by Chimakurthi *et al.,* (2009), Aono *et al.,* (2009) and Gordnier *et al.,* (2010). Chimakurthi *et al.,* (2009) found that the spanwise flexibility was shown to have a favourable impact on the thrust generation. Furthermore, the tip displacement increased over the entire range of reduced frequencies within the range of parameters considered. Aono *et al.,* (2009) found that the phase difference between the foil tips is a critical factor for thrust generation. The increment of the effective angle of attack at the foil tip due to the flexible motion is shown to be a secondary factor in the increase of thrust. A complex interaction between the vortices and the structural motion was observed by Gordnier *et al.,* (2010) with Large Eddy Simulations (LES). They found a stronger leading-edge vortex due to the higher effective angles of attack on the flexible foil. They also found regions of higher and lower pressure on the foil's upper and lower surfaces due to the larger accelerations of the flexible foil compared to the rigid foil. Kang *et al.,* (2011) considered both chordwise flexibility and spanwise flexibility and provided a generalised scaling function on the relationship between the thrust and maximum relative foil tip deformation as well as the propulsion efficiency. These studies on the spanwise flexibility suggested that the mean thrust coefficient is a function of the Strouhal number based on the prescribed heaving motion, and is very weakly dependent on the Reynolds number (Shyy *et al.,* 2010, Gursul *et al.,* 2014).

The propulsion performance of flexible propulsors at the resonance condition has been widely studied on membranes and thin wings, in which the cross-section of these flexible propulsors is rectangular with a large aspect ratio (Michelin and Llewellyn Smith 2009, Vanella *et al.,* 2009, Spagnolie *et al.,* 2010, Thiria and Godoy-Diana 2010, Yin and Luo 2010, Ramananarivo *et al.,* 2011). Kang *et al.,* (2011) provided the thrust scaling function for flapping flexible wings in water in forward motion without any structural or fluid damping considered, which shows a maximum thrust at the resonance frequency. They also suggested that the optimal propulsion performance will occur slightly below the natural frequency, due to the structural and fluid damping, consistent with findings from other studies.

Recently, Martínez-Muriel *et al.,* (2023) conducted direct numerical simulations of a spanwise-flexible wing in forward flight undergoes a combined heave and pitch motion. Linearised structural model was used. Low density fluid (air) was considered, which belongs to inertia-driven regime. Moderate Reynolds number (1000) was employed. The optimal aerodynamic performance can be achieved when the wing's oscillation frequency approaches the its natural frequency, in which the time averaged thrust reached the maximum. However, further increase the flexibility from the resonance point will results in a dramatic decrease of the thrust production as well as the propulsive efficiency. They found that the aspect ratio of the wing has limited effect on its aerodynamic performance. They found that the resonance phenomenon enhances aerodynamic performance by increasing effective angles of attack and delaying the development of the leading edge vortex.

Although Kang *et al.,* (2011) provided a generalised scaling function on the relationship between the thrust and maximum relative foil-tip deformation as well as the propulsion efficiency, and Martínez-Muriel *et al.,* (2023) shown the propulsive performance impacted by resonance effect under the inertia-driven regime, the impact of spanwise flexibility on the flow structure under the fluid-driven regime and its influence on the heaving foils propulsion is still unclear, especially when the natural frequency of the flexible foil varies over a wide range. Furthermore, the interaction between tandem heaving foils with spanwise flexibility, and the impact of the flexibility in this configuration on propulsion efficiency, is poorly understood. This work aims to address these two gaps. More specifically, this work employs numerical simulations to carry out decoupled analysis by retaining the density of the foil structure but varying the natural frequency (i.e. by changing the Young's modulus) of the foil structure, to investigate the impact of flexibility on the propulsion performance dual heaving foils arranged in tandem. This independent analysis is impossible to achieve experimentally. This study will provide insight into the physics of the problem of real-world materials, where discrete combinations of density and Young's modulus can be realised. The impact of in-phase motions will be studied since this provides a fundamental baseline configuration which can be extended in future studies. A wide range of natural frequencies, covering pre-resonance, resonance, and post-resonance regimes, with respect to the heaving frequencies, are studied. The flow structure, fluid forces acting on the foils, as well as the flexible motions of the foils, help to reveal the mechanism behind the impact of flexibility.

This paper is organised as follows: The definition of the problem, as well as the numerical setup, will be presented in Section 2. Results on the propulsion performance of rigid foil, including both rigid dual heaving foils and rigid single heaving foil, will be presented in Section 3.1. These results provide a benchmark for the propulsion performance analysis of flexible dual heaving foils in the following sections. Section 0 is the main section, in which the impact of the reduced frequency (i.e. the ratio between the heaving frequency and the natural frequency) of the flexible dual heaving foils on their propulsion performance will be studied. Section 3.3 investigates how propulsion performance of flexible dual heaving foils is affected by Strouhal number, either with a fixed reduced frequency or with a fixed natural frequency. The former is also a decoupled analysis that the natural frequency of the flexible dual heaving foils needs to match with the heaving frequency for a fixed reduced frequency, while the latter is about the propulsion performance of the flexible dual foils in reality, i.e. with a specific material at different working conditions. Conclusions are drawn in Section 4. Supplementary material is provided for a detailed analysis of the mechanism of rigid foil propulsion that is discussed in Section 3.1.

## 2. Problem Formulation and Numerical Methodology

### 2.1. Problem Description and Study Parameters

Two flexible foils in a tandem configuration are immersed in a fluid with an inlet flow speed, $u$, as shown in Figure 1 (a). The foils are identical in shape, structural properties, and the prescribed heaving motion. The foils have rectangular planforms with a chord length, $c = 0.1\ m$ and the aspect ratio is three (Heathcote et al., 2008). The teardrop shape has been widely used in studies of heaving flexible foils due to its low drag characteristics (Heathcote and Gursul 2007, Shyy et al., 2010, Wu et al., 2020). In this study, the cross-section of the foil is based on a classic teardrop shape, with a rounded nose at the front, but its rear is modified from a sharp taper into a trapezoidal shape with a gentle slope. The foil cross-section follows a piecewise function as

$$y = \begin{cases} \frac{5c}{12} g(x), & x \in \left[0, \frac{c}{10}\right) \\ \frac{5}{108} \left\{ 10x \left[g(0) - g\left(\frac{c}{10}\right)\right] + c \left[10g\left(\frac{c}{10}\right) - g(0)\right] \right\}, & x \in \left[\frac{c}{10}, c\right] \end{cases}, \quad (1)$$

where, $g(x)$ is the teardrop basis function defined as

$$g(x) = \sin\left[2\cos^{-1}\left(\frac{x}{c} + \frac{3}{250}\right)\right] \sin^{100}\left[\cos^{-1}\left(\frac{x}{c} + \frac{3}{250}\right)\right]. \quad (2)$$

The purpose of this modification is to retain the low drag of the foil while reducing the likelihood of flow separation and turbulence, thus ensuring that the observed flow structure is in the laminar regime. However, it is important to note that this modification comes at a cost, as the lift generated by the foil will be very small. The gap between the trailing edge or the fore foil and the leading edge of the hind foil is initialised at $1c$. The prescribed heaving motion of the dual foils follows a sinusoidal function as

$$y_d(t) = a_0 \cos(2\pi f_h t), \quad (3)$$

where $a_0$ is the heaving amplitude, $f_h$ is the heaving frequency, and $t$ is the time. The Reynolds number ($Re$) is defined as

$$Re \equiv \frac{\rho_f u c}{\mu} = 100, \tag{4}$$

where $\rho_f$ is the fluid density and $\mu$ is the dynamic viscosity. The reduced heaving amplitude ($A_h$) and the mass ratio ($M$) are defined as

$$A_h \equiv \frac{a_0}{c} = 0.2 \tag{5}$$

and

$$M \equiv \frac{\rho_s d}{\rho_f s} = 0.6, \tag{6}$$

where $\rho_s$ is the foil's density, $d$ is the chordwise averaged thickness of the foil, and $s$ is the span length of the foil. Values of $Re$, $A_h$ and $M$ are selected based on Zhang *et al.*, (2020), which are close to the characteristics of propulsors in nature, such as a fish tail. The Strouhal number, with respect to the foil's heaving motion, is defined as

$$St = \frac{2 f_h a_0}{u}. \tag{7}$$

The thrust coefficient is defined as

$$Ct = \frac{-drag}{\frac{1}{2}\rho_f u^2 c s}. \tag{8}$$

The dimensionless reduced frequency is given by

$$f^* = \frac{f_h}{f_n}. \tag{9}$$

The vertical deformation of the foil at the free tip in the $y$-direction with body-fitted coordinates $x_b y_b z_b$ is denoted by $\boldsymbol{D_{T_y}}$ which is non-dimensionalised by the chord length is

$$\widetilde{\boldsymbol{D_{T_y}}} = \frac{\boldsymbol{D_{T_y}}}{c}. \tag{10}$$

The maximum $\widetilde{\boldsymbol{D_{T_y}}}$ over a heaving cycle is defined as $\langle \widetilde{\boldsymbol{D_{T_y}}} \rangle_{max}$. The power coefficient of the foil is defined as

$$Cp = \frac{P}{\frac{1}{2}\rho_f u^3 c s}, \tag{11}$$

where $P$ is the power consumed by the foil to drive its heaving motion and is calculated as

$$P = -lift \times \frac{dy_d}{dt}. \tag{12}$$

The propulsion efficiency of a heaving foil is calculated by the ratio of cycle averaged thrust coefficient, defined as $\overline{Ct}$, to the cycle averaged power coefficient, $\overline{Cp}$, over a single heaving cycle as

$$\eta = \frac{\overline{Ct}}{\overline{Cp}}. \tag{13}$$

## 2.2. Numerical Method

In this study, we employ a parallel partitioned FSI framework (Liu *et al.,* 2020, Liu *et al.,* 2021, Liu *et al.,* 2022). The partitioned framework comprises three elements, namely the computational fluid dynamics (CFD) software, the computational structural mechanics (CSM) software, and the interface coupling library. In the present study, the CFD software is OpenFOAM-v6 (Weller *et al.,* 1998), the CSM software is FEniCS-v2019.1.0 (Alnæs *et al.,* 2015), while coupling is through the multiscale universal interface MUI-v1.0 (Tang *et al.,* 2015) library.

### 2.2.1. Fluid Solver

The governing equations are the incompressible Navier–Stokes equations. The continuity equation reads as

$$\nabla \cdot \boldsymbol{U} = 0, \tag{14}$$

where $\boldsymbol{U}$ is the fluid velocity vector. The momentum equation over an arbitrarily moving object, with structural deformation, is expressed as (Tuković *et al.,* 2018)

$$\frac{\partial \rho_f \boldsymbol{U}}{\partial t} + \nabla \cdot [\boldsymbol{U} \cdot \rho_f (\boldsymbol{U} - \boldsymbol{U}_s)] = -\nabla p + \nabla \cdot (\mu \nabla \boldsymbol{U}) + \boldsymbol{s}_\phi, \tag{15}$$

where $\boldsymbol{U}_s$ is the grid velocity, $p$ is the fluid pressure, and $\boldsymbol{s}_\phi$ is a source term. The space conservation law of an arbitrary Lagrangian-Eulerian (ALE) formulation that is used to close the momentum equation with an arbitrarily moving object under the Finite Volume method is defined as

$$\frac{d}{dt} \int_V dV - \oint_S \boldsymbol{n}_s \cdot \boldsymbol{U}_s \, dS = 0, \tag{16}$$

where $\boldsymbol{n}_s$ is the unit normal vector pointing outward of surface $S$. Equations (14) and (15) are coupled via the Pressure-Implicit with Splitting of Operators (PISO) algorithm (Issa 1986). Second-order Gauss linear schemes are employed and a second-order Crank-Nicolson scheme is used for the temporal integration.

### 2.2.2. Structural Solver

The structural solver is based on the elastodynamics formulation expressed in the form of a generalised *n*-DOF (degrees of freedom) harmonic oscillator. In terms of body-fitted coordinates, the solid structure is given by

$$\boldsymbol{M} \frac{\partial^2 \boldsymbol{d}}{\partial t^2} + \boldsymbol{C} \frac{\partial \boldsymbol{d}}{\partial t} + \boldsymbol{K} \boldsymbol{d} = \boldsymbol{F}(t), \tag{17}$$

where , $\boldsymbol{C}$, and $\boldsymbol{K}$ are the mass, damping, and stiffness matrices, respectively, $\boldsymbol{F}$ is an external load, which is a function of time, and $\boldsymbol{d}$ is the deformation with respect to the body-fitted coordinates of the solid structure. In the present solver, the damping matrix is modelled based on Rayleigh damping as

$$\boldsymbol{C} = \alpha_m \boldsymbol{M} + \alpha_k \boldsymbol{K}, \tag{18}$$

where $\alpha_m$ and $\alpha_k$ are Rayleigh damping parameters. Combining Equations (17) and (18) give

$$\boldsymbol{M} \frac{\partial^2 \boldsymbol{d}}{\partial t^2} + (\alpha_m \boldsymbol{M} + \alpha_k \boldsymbol{K}) \frac{\partial \boldsymbol{d}}{\partial t} + \boldsymbol{K} \boldsymbol{d} = \boldsymbol{F}(t), \tag{19}$$

Equation (19) is discretised via the Finite Element method. Second-order quadratic Lagrange tetrahedral-based elements are employed. The generalised-$\alpha$ method, which is an extension of the Newmark-$\beta$ method (Newmark 1959), is used to achieve a second-order accuracy for the time stepping (Erlicher *et al.,* 2002, Bleyer 2018).

The generalised harmonic oscillator equation has been shown to have good accuracy on resonance prediction (Liu *et al.,* 2001, Wang *et al.,* 2005, Demirer *et al.,* 2021) and simulation of a flexible structure with tip displacement up to 0.25 times the foil span (Luo *et al.,* 2020) the largest foil tip displacement is 0.17 times the foil span.

2.2.3. Fluid-Structure Interaction

The fluid and structural domains of the present framework are coupled by kinematic and dynamic conditions at the interface. To ensure consistency, the displacement of the fluid-structure interface should adhere to a kinematic condition, while the fluid forces, or tractions, exerted on the interface must satisfy a dynamic condition for conservation. The kinematic condition reads (Slyngstad 2017)

$$\boldsymbol{d_s} = \boldsymbol{d_f}, \quad (20)$$

where $\boldsymbol{d_s}$ and $\boldsymbol{d_f}$ represent the displacement at the interface in the structural and fluid domains, respectively. The fluid forces follow the dynamic condition (Slyngstad 2017, Tuković *et al.,* 2018)

$$\sigma_s \cdot \boldsymbol{n_s} = \boldsymbol{t_f}, \quad (21)$$

where $\sigma_s$ is the stress tensor of the structural domain, and $\boldsymbol{t_f}$ is the traction at the interface, which is calculated as

$$\boldsymbol{t_f} = \sigma_f \cdot \boldsymbol{n_s}, \quad (22)$$

The stress tensor at the interface, $\sigma_f$, which is calculated from the fluid domain for an incompressible Newtonian fluid, is expressed as

$$\sigma_f = -p\boldsymbol{I} + \tau, \quad (23)$$

where $\tau$ is the viscous component of the stress tensor, calculated as

$$\tau = \mu_f(\nabla \boldsymbol{U} + \nabla \boldsymbol{U}^T), \quad (24)$$

the transpose of matrix $\boldsymbol{A}$ being denoted as $\boldsymbol{A}^T$. The dynamic condition dictates that the forces acting on the interface are conserved between the two domains. In the discretised form, this is achieved using the radial basis function (RBF) spatial interpolation method in the MUI library. The RBF spatial interpolation can handle general nonconformal meshes regardless of gaps between meshes or different mesh densities (Rendall and Allen 2008, Bungartz *et al.,* 2016). The interpolation from a source mesh to a target mesh is based on radial basis functions computed on each element/cell centre of the source mesh. Both Gaussian and Wendland's functions are available as the basis functions in MUI, while the Gaussian basis function is used in this study. Both the fluid and the structural solver perform a number of sub-iterations for the interaction at each time step. At any time step and sub-iteration, the fluid solver updates the fluid domain, and the fluid forces in each cell located at the interface between the fluid and structural domains are obtained. In this framework, the fluid forces at each cell of the interface are transferred to the structural domain using the message passing interface (MPI) through the MUI library. The structural domain fetches fluid forces and applies them as the right-hand-side of Eq. (19). It then updates the deformation of the structure and pushes this information back to the fluid solver, again using MUI. The displacements of the structure in each cell of the interface are transferred and applied to the fluid domain as a Dirichlet boundary condition. Both fluid and structural domains are advanced to the next sub-iteration after the completion of these steps. A number of sub-iterations are typically needed within each time step until the residual on the fluid and structural coupling is below a tolerance. Aitken's method (Chow and Kay 1984) is employed to achieve a tight coupling.

The parallel partitioned FSI framework, as well as the numerical algorithms applied herein, have been extensively verified and validated through systematic benchmark cases in our previous publications, covering both two-dimensional and three-dimensional aero/hydroelastic cases within both laminar and turbulent regimes (Liu *et al.*, 2020, Liu *et al.*, 2021, Liu *et al.*, 2022). Good agreement with published numerical and experimental results has been achieved, which demonstrates the accuracy of the framework. Importantly, we have been able to capture the resonance effect of an elastic hydrofoil accurately against published experimental data (Liu *et al.*, 2020, Liu *et al.*, 2021).

### 2.3. Geometry, Mesh and Boundary Conditions

The fluid domain extends $10c$ from the leading edge of the fore foil towards the upstream inlet, $25c$ away from the trailing edge of the hind foil towards the downstream outlet, $12.5c$ away from the upper surface of the foil towards the maximum *y*-axis boundary, $12.5c$ away from the lower surface of the foil towards the minimum *y*-axis boundary and $12.5c$ away from the foil's forced heaving tip and free tip towards the minimum and maximum *z*-axis domain boundaries, respectively. The size of the domain is determined from published studies (Sun *et al.*, 2018, Majumdar *et al.*, 2020, Thekkethil *et al.*, 2020, Zhang *et al.*, 2020, Zhang *et al.*, 2022) and our previous oscillating foil simulations (Liu *et al.*, 2020, Liu *et al.*, 2021), this to ensure the foils are sufficiently far from these boundaries to capture the important flow features and avoid numerical instabilities. The origin of the fluid computational domain is set at the leading edge of the fore foil at the side the displacement is prescribed, as shown in Figure 1 (a). A constant and uniformly distributed velocity along the *x*-axis direction, $u$, is applied to the inlet boundary along the *x*-axis direction as the incoming flow. The pressure gradient of the inlet boundary and the velocity gradient of the outlet boundary are zero. The absolute pressure of the outlet boundary is set to be the reference pressure. The minimum and maximum boundaries in the *z*-axis and *y*-axis directions of the fluid domain are modelled as symmetry planes, enforcing the gradient of the velocity and pressure in the direction normal to the boundary to be zero. A no-slip wall is applied to the dual foils. The structural domain is shown in Figure 1 (b), the dual foils have a forced heaving tip at $z = -\frac{3c}{2}$ plane and a free tip at $z = \frac{3c}{2}$ plane. The motion of the forced heaving tip is constrained by the heaving function and the motion of the free tip is fully solved by the elastic functions. Following a grid-independent study, a 3-D block-structured grid with 19 million hexahedral cells is adopted for the fluid domain and a 3-D tetrahedral grid with 7 thousand degrees of freedoms is adopted for the structural domain. As can be seen from the foil cross-section formulation, the foil has blunted leading and trailing edges with about $0.01c$, which is for the convenience of mesh generation for the dual foils. The classic H-type pattern is employed for the meshing of the foil in the fluid domain. Both the fluid solver and the structural solver in the present framework have the same time step size (Liu *et al.*, 2022). The time step size is adopted after carrying out a convergence study, to ensure that the Courant number in the fluid domain is less than 0.5.

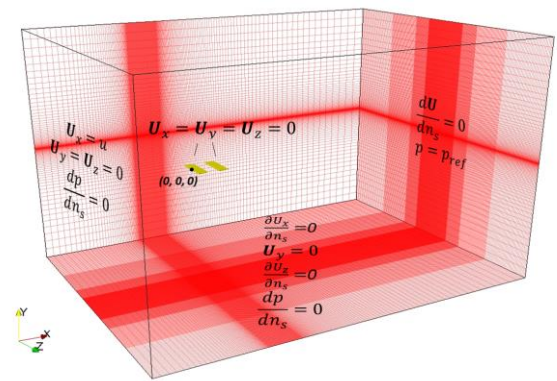

(a) Fluid domain

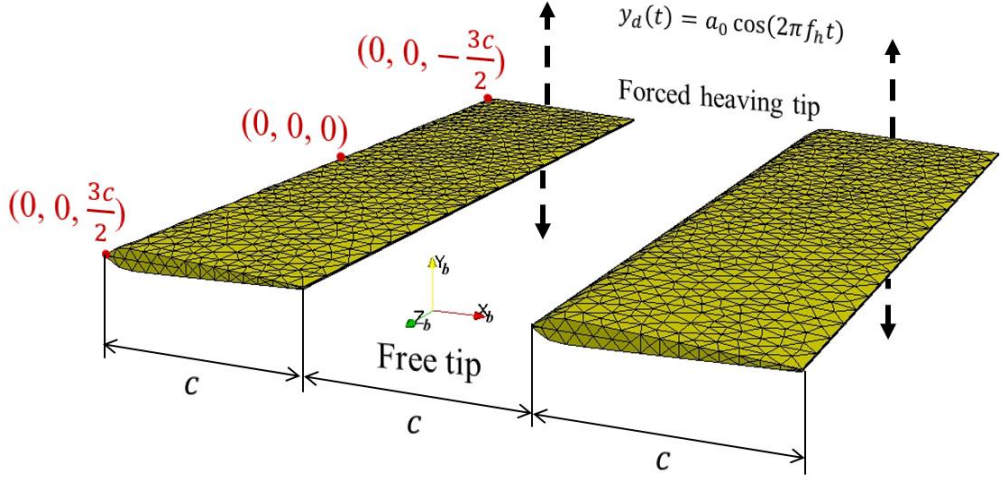

(b) Structural domain

Figure 1. Mesh generation and boundary conditions for (a) fluid domain and (b) structural domain.

## 3. Results and Discussion

This section begins with the study of rigid dual heaving foils, which will act as reference results for studying flexible dual heaving foils in the latter part of this section. This is followed by the study of the reduced frequency effect of the flexible dual heaving foils, in which we take advantage of numerical simulations by varying the natural frequency of the flexible foils and investigating its impact on the propulsion efficiency and its mechanism. We also investigate the propulsion performance of the flexible dual heaving foils under different Strouhal numbers with either fixed reduced frequency or fixed natural frequency, as in real working conditions.

### 3.1. Benchmark Cases

A key parameter of interest is the propulsion efficiency, ($\eta$). Figure 2 shows the propulsion efficiency, as defined by Eq. (13), and the cycle averaged components over a heaving cycle of the fore, hind and dual foils compared with a single heaving foil for different Strouhal numbers. A peak propulsion efficiency is achieved around $St = 0.64$ for both single foil and dual foils under the present working conditions. The dual foils have a lower propulsion efficiency in comparison with the single heaving foil, as shown in Figure 2 (a). The peak propulsion efficiency of the rigid dual foils (around 3%) is relatively low due to the parameters of our study; we take where the foil dimension, shape, and structural properties are from Heathcote et al. (2008), Heathcote and Gursul (2007), and Zhang et al. (2020), respectively. In the small Strouhal number region ($St < 0.64$), big differences in the propulsion efficiency are observed. The hind foil has the lowest propulsion efficiency in this region, compared with that of the fore foil and the single foil. In the large Strouhal number region ($St > 0.64$), the propulsion efficiency of the foils approaches a similar value, with a difference smaller than 5%.

Since a direct comparison between the dual and single foils of the thrust and power coefficients is not appropriate because of the different surface area involved, we further consider the idealised case where the fore and hind foils are isolated from one another due to infinite spacing. The thrust coefficients and the power coefficients of both the fore foil and hind foil are equal to that of the single foil (and can therefore be computed via a single foil simulation).

The cycle averaged thrust and power coefficients of each foil are compared with those of a single foil. Our results show that the fore foil generates almost the same thrust as the single foil, but requires slightly more power. In contrast, the hind foil generates slightly less thrust than the single foil but requires slightly less power for all Strouhal numbers considered. These results show that the main factor affecting the propulsion efficiency loss of the dual foils in the small Strouhal number region, as is observed in Figure 2 (a), is the thrust loss of the hind foil due to the interaction between the fore and aft foil.

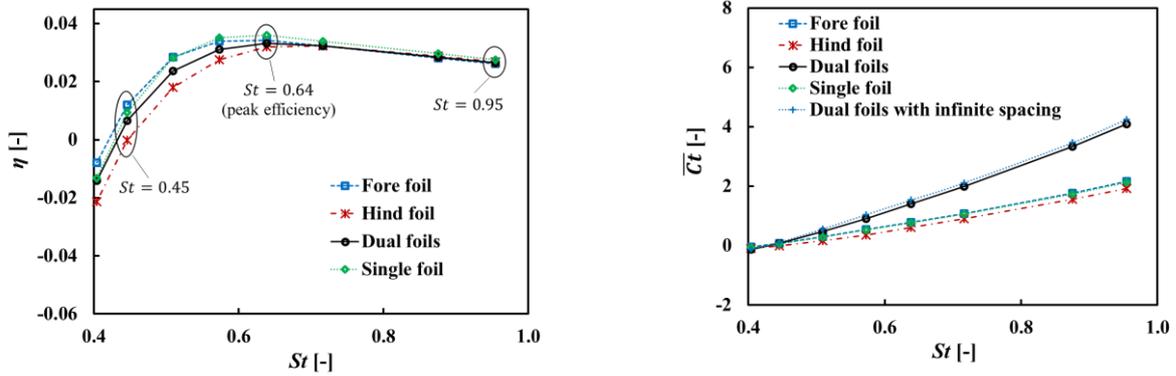

(a) Propulsion efficiency        (b) Thrust coefficient

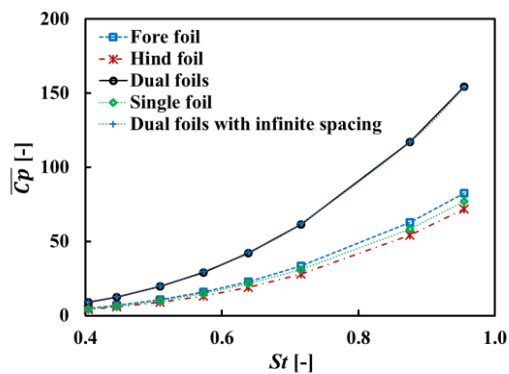

(c) Power coefficient

Figure 2. Propulsion performance of rigid single heaving foil and rigid dual heaving foils with (a) propulsion efficiency, (b) thrust coefficient and (c) power coefficient.

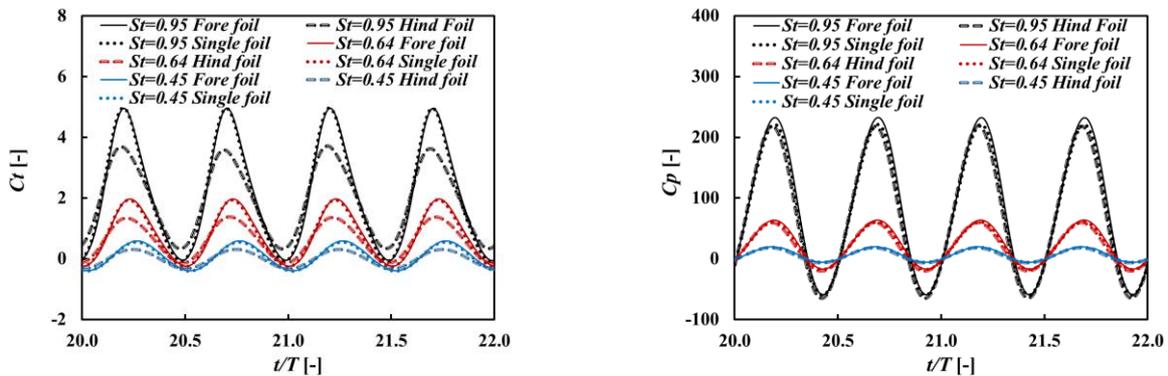

(a) Thrust coefficient        (b) Power coefficient

Figure 3 Instantaneous plots of rigid dual foils at three Strouhal numbers with (a) thrust coefficient and (b) power coefficient.

The instantaneous thrust and power coefficients of rigid dual foils at three Strouhal numbers have been plotted in Figure 3. For all three Strouhal numbers, the thrust coefficient of the fore foil is close to that of the single foil, while the thrust coefficient variance is less for the hind foil (Figure 3 (a)). The power coefficient of the single foil has a smaller amplitude than that of the fore foil but has a larger amplitude than that of the hind foil, as shown in Figure 3 (b). These observations are consistent with what we see in Figure 2 (b) and (c).

### *3.2. Flexible Dual Foils with Fixed Strouhal Number*

This section presents the propulsion performance of flexible dual heaving foils with different reduced frequencies at three Strouhal numbers. The reduced frequency is changed, with a fixed Strouhal number, i.e. the heaving frequency, $f_h$, remains constant. Under this condition, the natural frequency of the foil structure, $f_n$, is changing when $f^*$ changes. Since the square of the natural frequency is proportional to Young's modulus of an elastic material with a given shape and density (Spinner 1961, Heritage *et al.,* 1988), the change of $f_n$ is the change of the flexibility (i.e. Young's modulus) of the foil. It is a virtual analysis, which is hard to achieve through experiment as it is challenging to alter the natural frequency (or Young's modulus) of a flexible heaving foil continuously over a large interval. The reduced frequency of flexible dual heaving foils ranges from 0 to 1.5 within this study. The reduced frequency of the heaving foil equals 0, indicating $f_n$ tends to infinity, which represents a rigid heaving foil. When the reduced frequency of the heaving foil equals unity, $f_h$ and $f_n$ are equal, which corresponds to the resonance regime. For $f^* = 1$ as the distinguishing point, the pre-resonance regime is $0 < f^* < 1$ and the post-resonance regime $f^* > 1$.

    3.2.1. Propulsion Performance

The propulsion efficiency, thrust coefficient, power coefficient, and tip deformation coefficient of flexible dual heaving foils at three Strouhal numbers with reduced frequency from 0 to 1.5 are shown in Figure 4 Here, the propulsion efficiency, thrust force, and power consumption decrease when the dual heaving foils are close to the resonance regime (i.e. $f^*$ approaching 1). The tip deformation coefficient jumps to a maximum value in the resonance regime for all three Strouhal numbers, and its value decreases when $f^*$ keeps increasing, as shown in Figure 4 (d). When $St = 0.45$, the propulsion efficiency of the flexible dual foils almost quadruples when $f^*$ increases from 0 to 0.81. After the sudden drop at $f^* = 1$, the propulsion efficiency of the dual foils slightly rebound when $f^*$ exceeds 1 and continues to increase.
The propulsion efficiency of the flexible dual foils at $St = 0.64$ and $St = 0.95$ have different behaviour than that of the flexible dual foils at $St = 0.45$, which is quite stable with about 5% decrease at $St = 0.64$ and 9% decrease at $St = 0.95$, when $f^*$ increases from 0 to 0.81. The propulsion efficiency of the flexible dual foils also have a slight rebound when $f^*$ exceeds 1 and continues to increase at both $St = 0.64$ and $St = 0.95$.
In these three Strouhal numbers, both $\overline{Ct}$ and $\overline{Cp}$ increase with $f^*$ within $0 < f^* < 0.81$. When $f^*$ approaches to 1, both $\overline{Ct}$ and $\overline{Cp}$ drop rapidly and are lower than their values at $f^* = 0$. After $f^* = 1$ both $\overline{Ct}$ and $\overline{Cp}$ rise slightly.

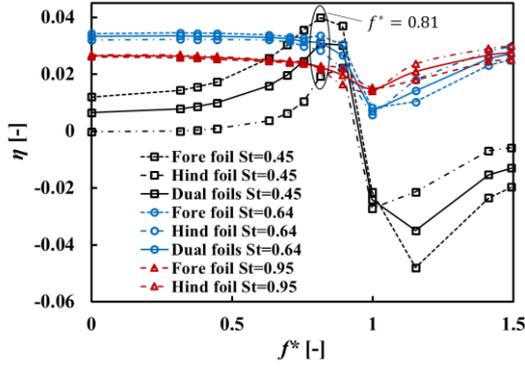
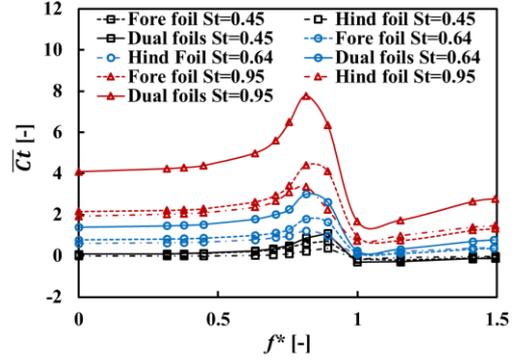

(a) Propulsion efficiency

(b) Thrust coefficient

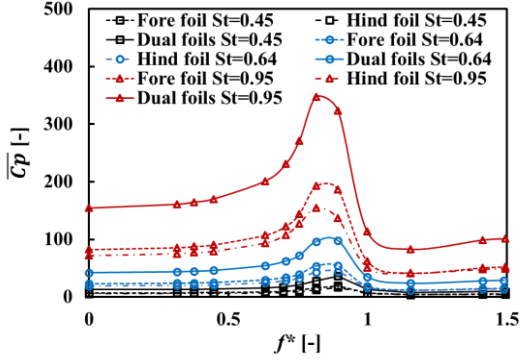
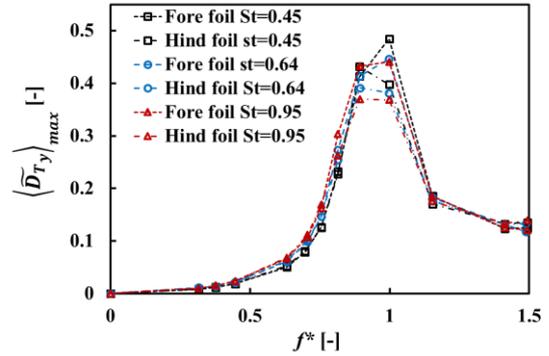

(c) Power coefficient

(d) Tip deformation coefficient

Figure 4. Propulsion performance of flexible dual heaving foils for $0 \leq f^* \leq 1.5$ for three Strouhal numbers with (a) propulsion efficiency, (b) thrust coefficient, (c) power coefficient and (d) tip deformation coefficient.

Figure 5 shows the rate of change (ROC) of $\overline{Ct}$, $\overline{Cp}$ and $\eta$ of flexible dual heaving foils for different $f^*$ with respect to rigid heaving foils at three Strouhal numbers. For all three Strouhal numbers, when $0 < f^* < 1$, the ROC of $\overline{Ct}$ and $\overline{Cp}$ increases with $f^*$ and reaches a peak around $f^*$ between 0.8 and 0.9. The ROC of $\overline{Ct}$ and $\overline{Cp}$ of flexible dual heaving foils with respect to rigid heaving foils reaches 1234% and 187%, respectively, at $f^* = 0.89$ with $St = 0.45$. The growth rate of $\overline{Ct}$ is an order of magnitude higher than that of $\overline{Cp}$, which leads to a 366% increase of the ROC of $\eta$ with respect to rigid heaving foils of the flexible foils at $f^* = 0.89$ with $St = 0.45$. When $St = 0.64$ and $St = 0.95$, the ROC of $\overline{Ct}$ with respect to rigid heaving foils is less than 120% and the ROC of $\overline{Cp}$ with respect to rigid heaving foils is above 125%, which leads to a negative ROC of $\eta$ with respect to rigid heaving foils between $f^* = 0.8$ and $f^* = 0.9$. Where $f^*$ approaches 1, the ROC of $\overline{Ct}$ and $\overline{Cp}$ drop to negative values (i.e. the cycle averaged thrust and the cycle averaged power coefficient of these foils at $f^* = 1$ are lower than that of rigid dual heaving foils) for all three Strouhal numbers. Even more, the ROC of $\overline{Ct}$ drops more than the ROC of $\overline{Cp}$ when $f^*$ approaches to 1. Therefore, the propulsion efficiency of flexible dual heaving foils at $f^* = 1$ is less than that of rigid dual heaving foils for all three Strouhal numbers.

In this subsection, our analysis of propulsion performance demonstrates that flexible foils hold the potential to enhance thrust to provide extra acceleration in comparison to rigid foils. Replacing rigid foils with flexible ones exhibits the potential to improve propulsion efficiency within the pre-peak propulsion efficiency regime. These thrust and propulsion efficiency enhancements, however, require additional power input. Conversely, resonance conditions prove to be the most unfavourable scenario, resulting in reduced thrust and diminished propulsion efficiency across all regimes when compared to rigid foils under the current working conditions. Notably, within the pre-resonance regime, the fore foil outperforms the hind foil in terms of propulsion performance. However, in the post-resonance regime, this relationship is reversed, with the hind foil surpassing the fore foil in propulsion performance across all cases.

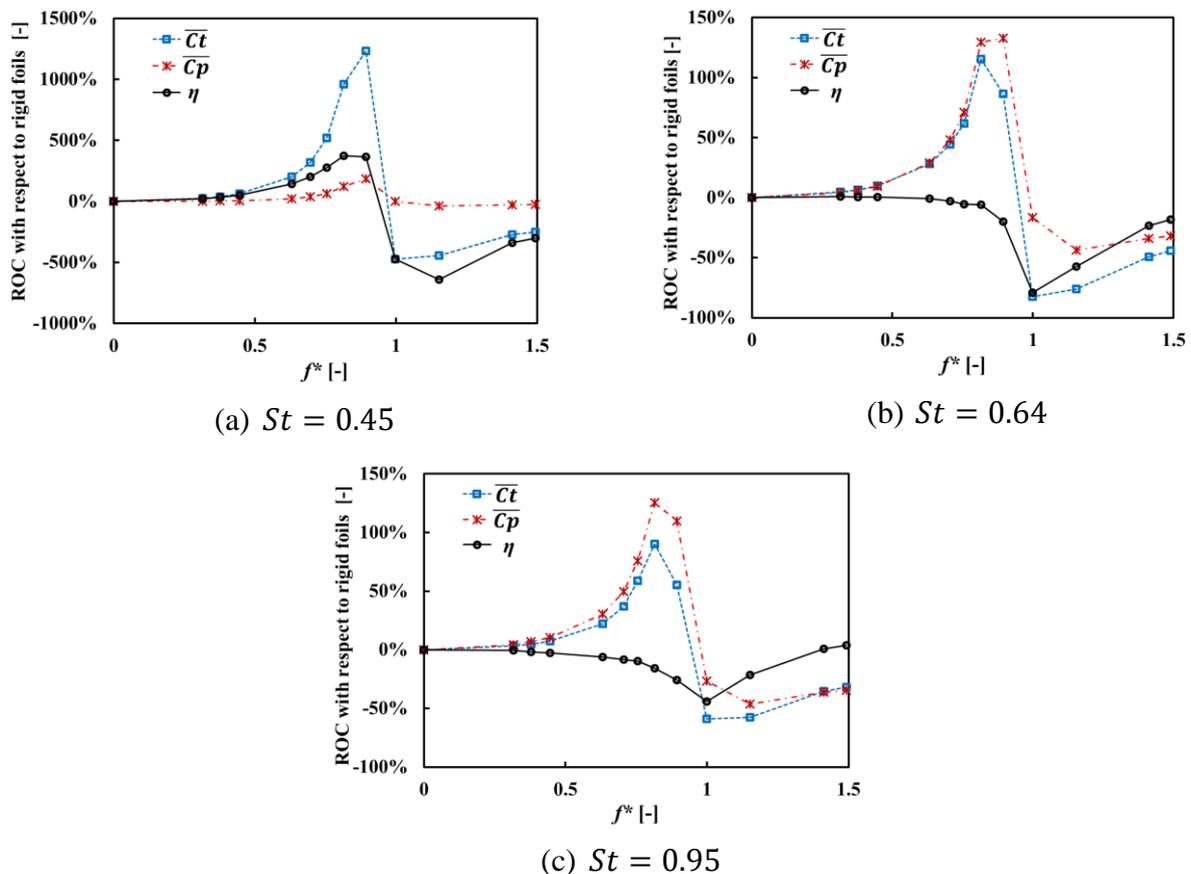

(a) $St = 0.45$

(b) $St = 0.64$

(c) $St = 0.95$

Figure 5. Rate of change of $\overline{Ct}$, $\overline{Cp}$ and $\eta$ compared to rigid heaving foils with (a) $St = 0.45$, (b) $St = 0.64$ and (c) $St = 0.95$.

3.2.2. Instantaneous Analysis on the Propulsion Performance at $St = 0.45$

To gain insight into the propulsion performance shown in Section 3.2.1, particularly with regard to the instantaneous behaviour of these physical quantities during different cycles, Figure 6 shows the instantaneous thrust coefficient, power coefficient and tip deformation as well as the power spectral density of the tip deformation coefficient of flexible dual heaving foils at $St = 0.45$ with four reduced frequencies. The instantaneous $Ct$ and $Cp$ of dual foils have the largest peak values at $f^* = 0.81$ compared with other reduced frequencies. A 90° phase difference on the instantaneous $Ct$ between fore foil and hind foil at $f^* = 1$ is observed from Figure 6 (a), while no noticeable phase difference on the instantaneous $Cp$ at the same frequency of $f^* = 1$ can be observed between fore and hind foil from Figure 6 (b). By comparing Figure 6 (a) and (b) with Figure 3 (a) and (b), the impact of flexibility is dominant on the flexible fore foil at all three reduced frequency regimes since the difference in terms of the amplitude of both instantaneous $Ct$ and $Cp$ between the flexible fore foil and rigid fore foil is much larger than the difference between that of the rigid fore foil and the rigid single foil. This difference between the flexible fore and hind foil indicates that both the flexibility and the interaction between foils have a great impact on the flexible hind foil. From Figure 6 (c), we observe that $\widetilde{D}_{T_y}$ shows irregular behaviour from cycle to cycle on both the fore and hind foil at $f^* = 1.49$. With the plot of power spectral density of the tip deformation coefficient in

Figure 6 (d), two dominant frequencies appear on the tip deformation of the heaving foils at $f^* = 1.49$, one is the natural frequency of the foil and the other is the foil heaving frequency.

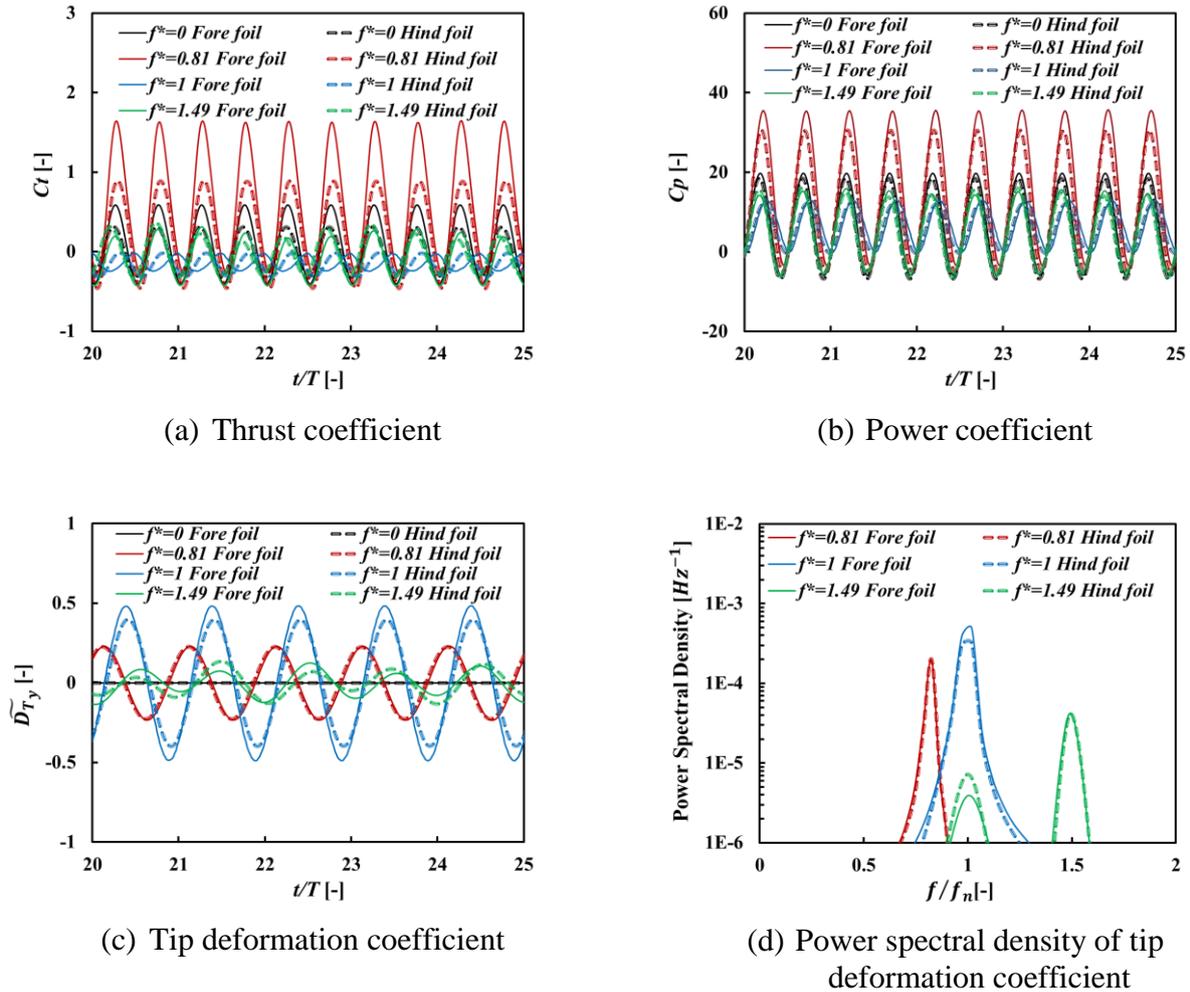

(a) Thrust coefficient

(b) Power coefficient

(c) Tip deformation coefficient

(d) Power spectral density of tip deformation coefficient

Figure 6. Instantaneous plots of flexible dual heaving foils at three reduced frequencies at $St = 0.45$ with (a) thrust coefficient, (b) power coefficient, (c) tip deformation coefficient and (d) power spectral density of tip deformation coefficient.

3.2.3. Elastic Behaviour at $St = 0.45$

The normalised mid-chord position of the flexible dual heaving foils ($\tilde{y}_{mid} = y_{mid}/c$) at three reduced frequencies at $St = 0.45$ is presented in Figure 7. The $x$-axis of Figure 7 represents the normalised position along the $z$-axis with a $\frac{3c}{2}$ offset towards the positive direction ($\tilde{z} = \frac{z}{c} + \frac{3}{2}$). The green dotted lines in Figure 7 represent the envelope of the deformation of the foil's mid-chord during one cycle. It is also an indication of the sectional peak-to-peak amplitude of the heaving foil along the spanwise direction. The sectional amplitudes of both the fore foil and hind foil gradually increase from the forced heaving tip to the free tip along the spanwise direction at $f^* = 0.81$. When $f^* = 1$, the sectional amplitudes of both foils decrease from the forced heaving tip towards the mid-span, then rise after passing to the mid-span until the free tip. Moreover, the heaving amplitude of the free tip of the hind foil is about 30% smaller than that of the fore foil at $f^* = 1$. When $f^* = 1.49$, the sectional amplitude of both the fore foil and hind foil gradually decreases from the forced heaving tip to the free tip along the spanwise direction. Phase lags are also observed on the foil's heaving motion of the free tip compared with that of the forced heaving tip for $f^* = 0.81$ and $f^* = 1$. At $f^* = 0.81$, when the forced heaving tip of the fore foil has passed its highest heaving position and is on its down-stroke, the free tip has just reached its highest heaving position. A similar phenomenon can be seen at $f^* = 1$, but with a larger phase lag between the foil's forced heaving and free tips. We observe no phase lag on the heaving motion of the free tip compared with the forced heaving tip for $f^* = 1.49$. The phase difference between the forced heaving and free tips of the heaving foil observed in Figure 7 is consistent with observations from Figure 6 (c), in which the differences are shown, in the heaving amplitude of the free tip between the fore and hind foil at $f^* = 1$, as well as the instability of the heaving motion of the free tip at $f^* = 1.49$. It is noted that the instability of the heaving motion of the foil at $f^* = 1.49$ is only for the free tip's heaving amplitude, but its phase of the heaving motion remains the same from cycle to cycle.

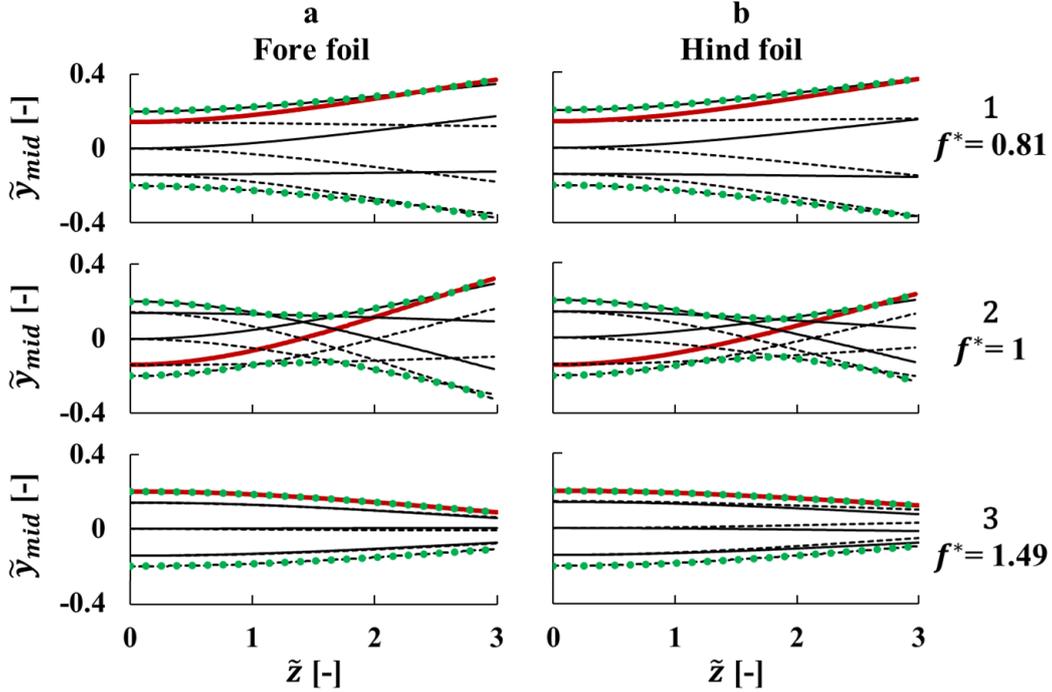

Figure 7 Mid-chord position of flexible dual heaving foils that is normalised by chord at three reduced frequencies at $St = 0.45$ with (a) fore foil, (b) hind foil, (1) $f^* = 0.81$, (2) $f^* = 1$ and (3) $f^* = 1.49$. The $x$-axis represents the normalised position along $z$-axis with a $\frac{3c}{2}$ offset towards the positive direction. Solid black lines represent the down-stroke motions. The dashed black line represents the up-stroke motions. The green dotted line represents the envelope of foil deformation. The red solid line represents the mid-chord position with the maximum deformation.

The effective angle of attack of a heaving foil is defined as

$$\alpha_{eff} = \alpha_p + \tan^{-1}\frac{dy_d}{u \cdot dt}, \quad (25)$$

where $\alpha_p$ is the geometric angle of attack of the heaving foil due to the pitch motion, which is zero in this study. The effective angle of attack of the heaving foil is a function of time and the maximum value of this effective angle of attack over a heaving cycle is defined as $\langle\alpha_{eff}\rangle_{max}$. Since $y_d$ is different along the spanwise direction for flexible foils, $\overline{\langle\alpha_{eff}\rangle_{max}}$ is used to define the spatial averaged maximum effective angle of attack. The reference effective angle of attack, $\langle\alpha_{eff}\rangle_{ref}$, is defined as the maximum effective angle of attack over a heaving cycle for a rigid single heaving foil when it reaches its peak propulsion efficiency. Figure 2 (b) and (c) show that both $\overline{Ct}$ and $\overline{Cp}$ continue to grow with increasing Strouhal number (i.e. raising the

$\overline{\langle\alpha_{eff}\rangle_{max}}$). When $\frac{\overline{\langle\alpha_{eff}\rangle_{max}}}{\langle\alpha_{eff}\rangle_{ref}} < 1$ and increasing $\overline{\langle\alpha_{eff}\rangle_{max}}$, the growth rate of $\overline{Ct}$ is greater than that of $\overline{Cp}$, which leads to an increase in the propulsion efficiency of the heaving foil. Conversely, when $\frac{\overline{\langle\alpha_{eff}\rangle_{max}}}{\langle\alpha_{eff}\rangle_{ref}} > 1$, the growth rate of $\overline{Ct}$ becomes smaller than that of $\overline{Cp}$ when increasing $\overline{\langle\alpha_{eff}\rangle_{max}}$, which leads to a decrease in the propulsion efficiency of the heaving foil. For $\frac{\overline{\langle\alpha_{eff}\rangle_{max}}}{\langle\alpha_{eff}\rangle_{ref}} = 1$, the peak propulsion efficiency of the heaving foil is reached.

Figure 8 (a) shows the spatial averaged maximum effective angle of attack relative to the $\langle\alpha_{\text{eff}}\rangle_{\text{ref}}$. The normalised maximum effective angle of attack of the flexible foils at $f^* = 0.81$ is larger than that of the rigid foils and close to unity. When the flexible foils are at $f^* = 1$ and $f^* = 1.49$, respectively, their normalised maximum effective angle of attack are less than that of the rigid foils. Figure 8 (b) shows the phase difference between the forced heaving tip and the free tip of the heaving foil. A phase difference of about $120°$ between the forced heaving tip and the free tip of the flexible heaving foils at $f^* = 1$ is observed. At $f^* = 0.81$, the phase difference between the forced heaving tip and the free tip of the heaving foils is about $30°$. The phase difference between the forced heaving tip and the free tip of the dual heaving foils at $f^* = 1.49$ is less than $10°$.

To enhance propulsion efficiency within the pre-peak propulsion efficiency regime, increasing the heaving amplitude through elastic displacement is an effective strategy for shifting $\alpha_{eff}$ towards the optimal range (i.e. $\langle\alpha_{eff}\rangle_{ref}$), thus resulting in additional propulsion efficiency gains. For instance, when examining the case of $f^* = 0.81$, flexible heaving foils outperform the rigid heaving foils due to the elastic displacement of the flexible material leads to a gradually enlarged heaving amplitude from the forced heaving tip to the free tip as shown in Figure 7 (1) and moves $\alpha_{eff}$ towards $\langle\alpha_{eff}\rangle_{ref}$.

However, it is important to note that a large elastic displacement does not always equate to higher propulsion efficiency. Take the example of $f^* = 1$, which has a similar scale of large elastic displacement with $f^* = 0.81$ as shown in Figure 4 (d). The key distinction lies in the large phase difference between the forced heaving tip and the free tip, leading to an almost inverted heaving motion at the free tip and a resulting reduction in heaving amplitude near the mid-span as shown in Figure 7 (2). This shift in motion causes $\alpha_{eff}$ to deviate from the optimal range and reduce propulsion efficiency.

Furthermore, elastic displacement does not necessarily translate to propulsion efficiency gains, even though there is only a minor phase difference between the forced heaving tip and the free tip. Consider the case of $f^* = 1.49$, the elastic displacement diminishes the heaving amplitude gradually from the forced heaving tip to the free tip, moving $\alpha_{eff}$ away from the optimal range. In summary, for a heaving foil operating in the pre-peak propulsion efficiency regime, the key to enhancing propulsion efficiency using flexible materials lies in ensuring that the elastic displacement enlarges the heaving amplitude, thus aligning $\alpha_{eff}$ with the optimal range. There is still a mystery as to why $f^* = 1$ has a slightly lower propulsion efficiency compared with that of $f^* = 1.49$ as to Figure 5 (a), but the conclusion drawn from Figure 8 (a) alone should be that $f^* = 1$ has a slightly better propulsion efficiency, which is the opposite. The following subsection will address this question and show the impact of the phase difference between the forced heaving tip and the free tip by examining the flow structure of dual heaving foils.

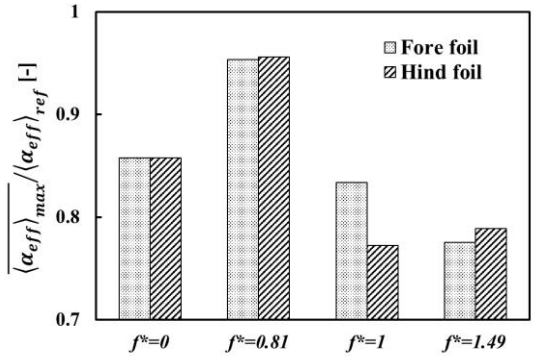
(a) Normalised maximum effective angle of attack

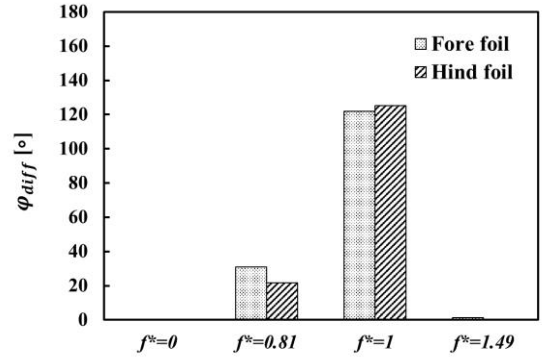
(b) The phase difference between the forced heaving tip and the free tip

Figure 8 Normalised maximum effective angle of attack (a) and phase difference between the forced heaving tip and the free tip (b) at four reduced frequencies at $St = 0.45$.

The chordwise angle of attack on the free tip of flexible foils is also plotted. It is induced by the torsion of the flexible foil, as shown in Figure 9. The largest maximum angle of attack at $St = 0.45$ is about $1.2°$ when $f^* = 1$. It is evident that the bending modes play a predominant role in shaping the structural behaviour under the present working conditions.

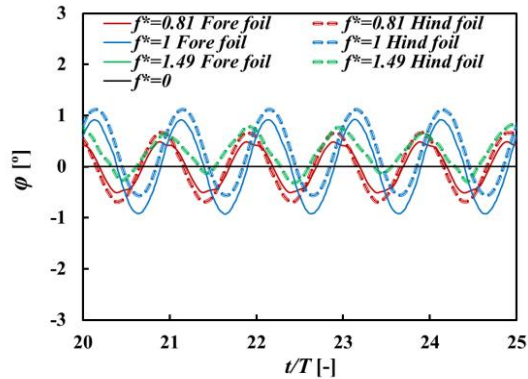

Figure 9 Angle of attack on the free tip of flexible foils under body-fitted coordinate at $St = 0.45$.

### 3.2.4. Three-Dimensional Flow Structure and Surface Contour at $St = 0.45$

The flow structure of the flexible dual heaving foils at $St = 0.45$ is investigated in this section to study the influence of the elastic motion, as observed and discussed in Section 3.2.3, on the flow field and its subsequent impact on propulsion performance. Additionally, we will investigate the interaction between the fore foil and the hind foil. As discussed in Section 3.2.2 that the impact of flexibility is dominant on the flexible fore foil but both the flexibility and the interactions between foils have a great impact on the flexible hind foil. In this section, the study of the flow structure on the flexible fore foil compared with that of the rigid fore foil is an effective way to figure out how it is influenced by the flexibility of foil, while the study of the flow structure on the flexible hind foil compared with that of the flexible fore foil is an effective way to figure out how it influenced by the interactions between the flexible dual foils, hence its influence on their individual propulsion performance.

Figure 10 shows the iso-surface of $Q$-criterion on dual heaving foils at $St = 0.45$ with different reduced frequencies and time instances. The TEV of the fore foil and LEV of the hind foil are symmetrical about the mid-span of the foil when $f^* = 0$. With the spanwise flexible motion at $f^* = 0.81$, the TEV of the fore foil is not symmetric about the mid-span. Instead, the TEV of the fore foil stretched on the side close to the forced heaving tip and squeezed on the side that is close to the free tip. The TEV of the hind foil is not symmetric about the mid-span as well. When $f^* = 1$, the TEV of the fore foil is further squeezed towards the free tip compared with that at $f^* = 0.81$. While at $f^* = 1.49$, we cannot observe the TEV of the fore foil being squeezed.

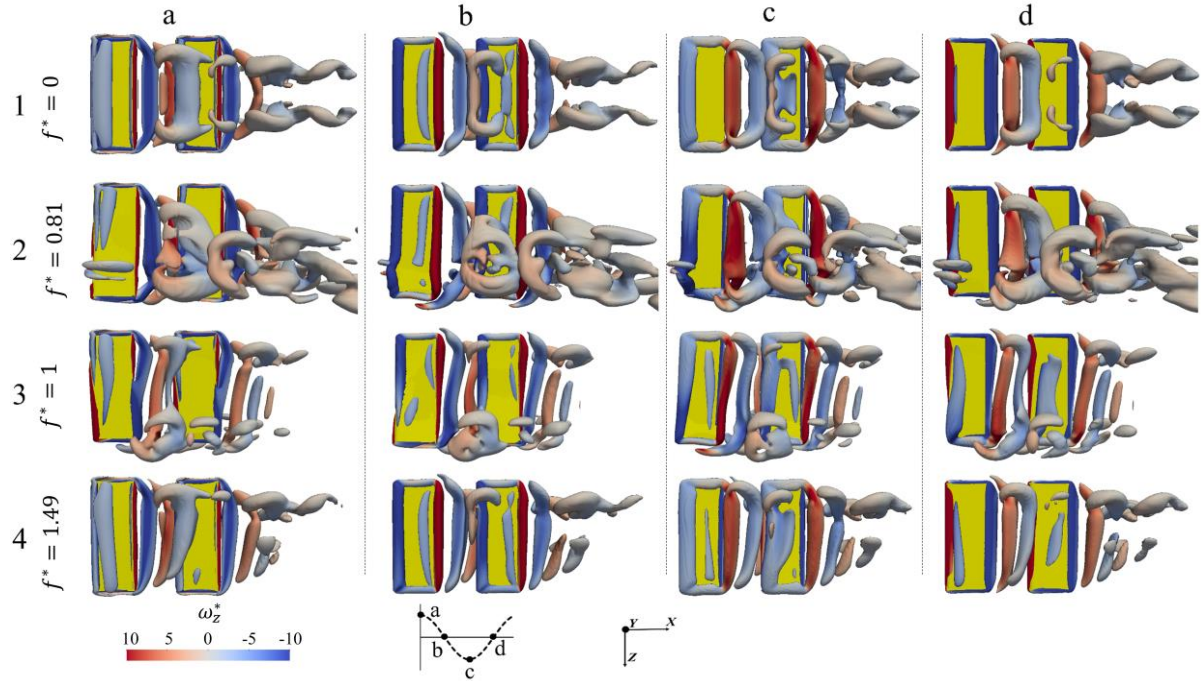

Figure 10 Iso-surface of $Q$-criterion ($\frac{Qc^2}{u^2} = 1$) coloured by the normalised $z$-vorticity for flexible dual heaving foils at $St = 0.45$ with (a) $t/T = 0$, (b) $t/T = 0.25$, (c) $t/T = 0.5$, (d) $t/T = 0.75$, (1) $f^* = 0$, (2) $f^* = 0.81$, (3) $f^* = 1$ and (4) $f^* = 1.49$.

A continuous phase shift of the LEV is observed from the forced heaving tip towards the free tip of the flexible foils. By comparing the LEV structure of the foils at $f^* = 1$ (Figure 10 (3)) and that of $f^* = 0$ (Figure 10 (1)), similar LEV structures at the forced heaving tip at each time instants are seen, but the LEV structure close to the free tip has a phase lag compared with the forced heaving tip one. Moreover, the heaving amplitude is different along the spanwise direction as indicated in Figure 7, which also leads to a different LEV structure. The differences in the LEV structure of the flexible foils are illustrated in Figure 10 (2)-(3) compared with that of the rigid foils illustrated in Figure 10 (1) are the combined effect of the phase difference between the forced heaving tip and the free tip as well as the different heaving amplitude along the foil spanwise direction. In addition, the phase lag on the LEV formation at $f^* = 1$ leads to the occurrence of two LEV with opposite sign of their vorticity at the same time instant at the leading edge of the foil. A reduction in the vorticity magnitude of the LEV of fore foil is observed at $f^* = 1$ around the forced heaving tip compared with that of the fore foil at $f^* = 0$. It corresponds to the impact between these two LEV with opposite vorticity sign. It should be noted that with the existence of the fore foil's TEV, the vorticity magnitude of the LEV of the hind foil around the forced heaving tip is larger than that of the fore foil at $f^* = 1$. The differences on the LEV structure of the foils at $f^* = 1.49$ compared with that at $f^* = 0$, as shown in Figure 10 (4), are purely dominated by the effect of heaving amplitude since the phase difference between the forced heaving tip and the free tip is zero.

We introduce a non-dimensional coefficient, named the facial thrust coefficient, which shows the distribution of the thrust/drag across the surface of the foil. The facial thrust coefficient of cell face $i$ in the fluid domain is defined as

$$Ct_i = \frac{-drag_i}{\frac{1}{2}\rho_f u^2 A_i}, \qquad (26)$$

where, $drag_i$ is the x-component of drag force acting on the $i$th cell face and $A_i$ is the area of the $i$th cell face, which includes the contribution from both the fluid pressure force and the fluid viscous force. Figure 11 shows the surface contour of the facial thrust coefficient of dual foils at $St = 0.45$ with different reduced frequencies and time instances. Comparing different reduced frequencies in Figure 11 shows a similar distribution of the facial thrust coefficient around the forced heaving tip of the foils at the same time instant but with different reduced frequencies. Significant differences are observed in the distribution of the facial thrust coefficient from the foil's mid-span to the free tip among different reduced frequencies. This is related to the observations in Figure 10 that the LEV structure of the foils are similar around the forced heaving tip but different around the mid-span and the free tip at each time instants. At $f^* = 0.81$, the sectional having amplitude of the foil along spanwise direction increases continuously with a 30° phase difference between the forced heaving tip and the free tip as shown in Figure 7. By comparing the fore foil's thrust distribution between the fore foil at $f^* = 0.81$ (Figure 11 (2)) and the fore foil at $f^* = 0$ (Figure 11 (1)), the fore foil at $f^* = 0.81$ generates more thrust (in terms of the magnitude and the area of the region where the thrust coefficient is positive) than that of the fore foil at $f^* = 0$ at all of the four time instants, especially around the free tip. The contribution of the 30° phase difference between the tips is to delay the thrust generation around the free tip of the foil by a one-twelfth cycle. While the larger area on the positive thrust coefficient around the free tip that we observed from Figure 11 (2a)-(2d) is dominated by the contribution of the larger sectional having amplitude (i.e. larger $\langle \alpha_{\text{eff}} \rangle_{\text{max}}$) around the free tip of the foil. It is consistent with what is observed from the instantaneous thrust coefficient in Figure 6 (a) that the thrust coefficient of the fore foil at $f^* = 0.81$ is larger than that of the fore foil at $f^* = 0$ throughout the cycle. At $f^* = 1$ Figure 11 (3) shows that the 120° phase difference between the tips dominates the thrust coefficient distribution on the fore foil's surface; a discontinuous distribution on the thrust coefficient between the half foil near the free tip and the half foil near the forced heaving tip can be observed. A phase shift of 120° leads to an opposite heaving direction of the foil's free tip compared with that of the forced heaving tip for most of a heaving cycle. It is reflected by the surface contour of the facial thrust coefficient in Figure 11 (3). At $t/T = 0$ and $t/T = 0.25$, half of the foil is thrust dominated, while the other half is drag dominated. At $t/T = 0.5$ and $t/T = 0.75$, although thrust is dominant, both halves of the foil, the magnitude of the thrust coefficient is very small around the foil's mid-span due to the small heaving amplitude (i.e. small $\langle \alpha_{eff} \rangle_{max}$) of that region. The reduction on the vorticity magnitude of the LEV of fore foil at $f^* = 1$ has a direct impact on the thrust generation at the leading edge of the fore foil around the forced heaving tip. Figure 11 (3) shows that the facial thrust coefficient of fore foil at $f^* = 1$ is smaller than that of the fore foil at $f^* = 0$ around the forced heaving tip. With the combined impact of these effects, the instantaneous thrust coefficient of the fore foil at $f^* = 1$ is flatter with a phase difference than that of the fore foil at $f^* = 0$ as shown in Figure 6 (a). At $f^* = 1.49$, the thrust distribution is non-symmetrical along the spanwise of the fore foil. It is caused by the small heaving amplitude and $\langle \alpha_{eff} \rangle_{max}$ around the free tip as presented in Figure 10. The magnitude of the facial thrust coefficient around the free tip of the fore foil is smaller than that of the forced heaving tip as can be seen from Figure 11 (4) and the instantaneous thrust coefficient of the fore foil at $f^* = 1.49$ is smaller than that of the fore foil at $f^* = 0$ throughout the cycle as shown in Figure 6 (a).

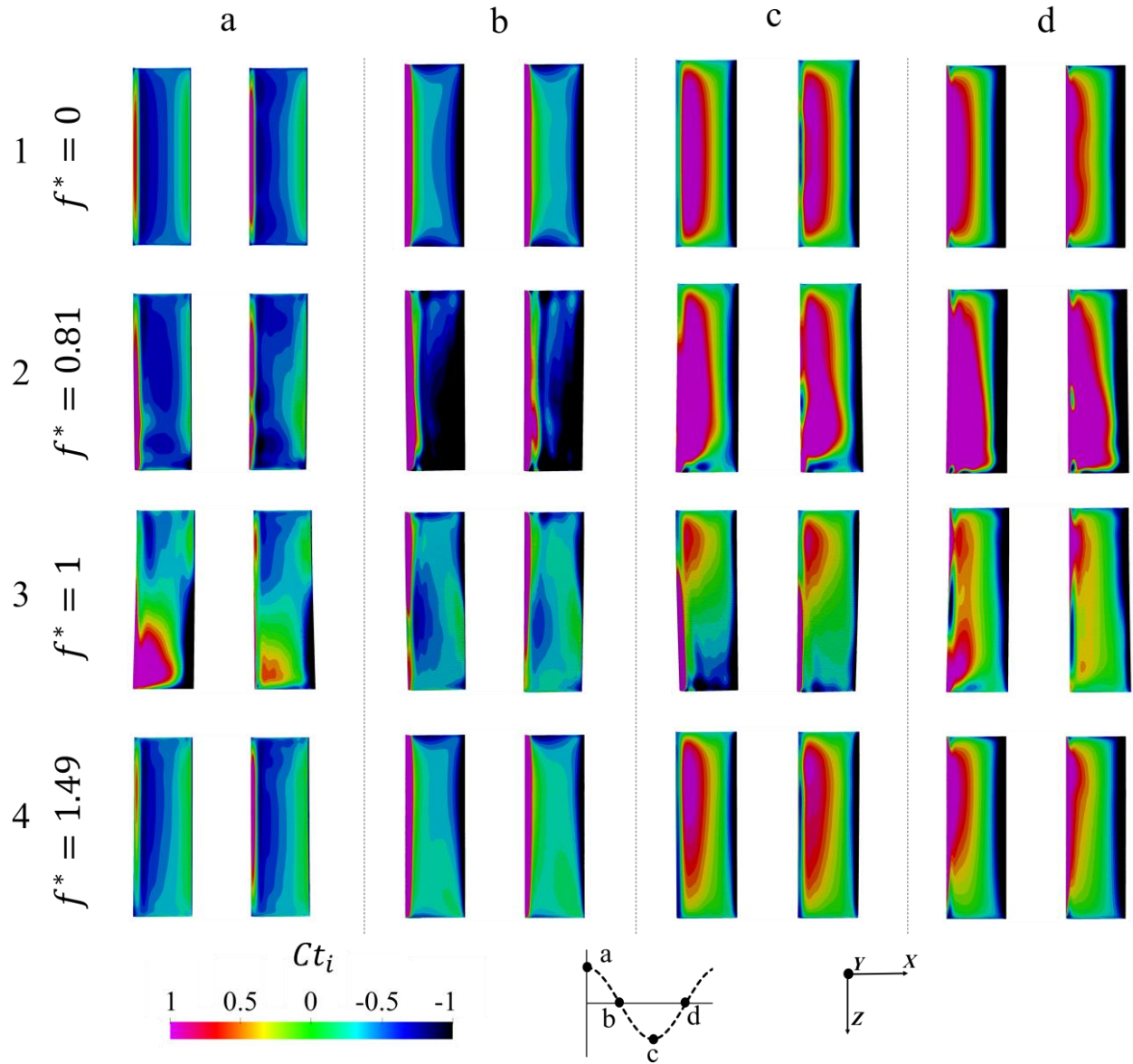

Figure 11 Surface contour of facial thrust coefficient at $St = 0.45$ with (a) $t/T = 0$, (b) $t/T = 0.25$, (c) $t/T = 0.5$, (d) $t/T = 0.75$, (1) $f^* = 0$, (2) $f^* = 0.81$, (3) $f^* = 1$ and (4) $f^* = 1.49$.

The vortex structure and the thrust distribution of the flexible hind foil are more complex than that of the flexible fore foil due to the interaction between the TEV from the fore foil and the LEV of the hind foil. From the conclusions drawn in the flow field analysis of rigid foil propulsion in the Supplementary Material, there will be a thrust enhancement at the leading edge of the hind foil when the LEV of the hind foil and the TEV of the fore foil have the same sign in terms of their vorticity. Otherwise, a thrust reduction at the hind foil's leading edge will occur when the LEV of the hind foil and TEV of the fore foil have the opposite sign in terms of their vorticity. It is also applicable to flexible foils. Figure 2 (a) in the Supplementary Material shows that the thrust enhancement effect occurs on the hind foil at $t/T = 0$. A similar effect can be observed in Figure 11 (a). A special focus is given to the flexible hind foil with $f^* = 1$ at $t/T = 0$ as in Figure 10 (a3) that the LEV generated by the half of the hind foil near the forced heaving tip has a negative sign in terms of its vorticity, while the LEV generated by the other half of the hind foil has a positive sign due to the phase difference on the heaving motion between the tips. When the hind foil's LEV interacted with the fore foil's TEV with a negative sign in terms of its vorticity, the half of the hind foil near the forced heaving tip will show thrust enhancement effect and the other half of the hind foil will show thrust reduction effect as can be observed from Figure 11 (a3). As in Figure 2 (b)-(d) in the Supplementary Material, the thrust reduction effect occurs on the hind foil due to the opposite sign of vorticity between the fore foil's TEV and hind foil's LEV from $t/T = 0.25$ to $t/T = 0.75$. The same phenomenon can be seen in Figure 10 (b)-(d) and Figure 11 (b)-(d). As the TEV of the fore foil stretched on the side close to the forced heaving tip and squeezed on the side that closed to the free tip for $f^* = 0.81$, the drag reduction effect also shifted from around the mid-span towards the free tip of the hind foil, as can be seen in Figure 11 (2b)-(2d). As observed from Figure 10 (3), the vorticity magnitude of the LEV of the hind foil around the forced heaving tip is larger than that of the fore foil at $f^* = 1$ due to the existence of the fore foil's TEV. Its impact can be seen in Figure 11 (3) where the magnitude of the facial thrust coefficient of the hind foil around the forced heaving tip is larger than that of the fore foil at $f^* = 1$. The thrust reduction phenomenon around the free tip of the hind foil at $t/T = 0.5$ and $t/T = 0.75$ and drag reduction phenomenon around the free tip of the hind foil at $t/T = 0.25$ are presented in Figure 11 (3b)-(3d). At $f^* = 1.45$, where the interaction between the fore foil's TEV and the hind foil's LEV has a larger impact on the thrust generation of the hind foil. This leads to a reduction in the facial thrust coefficient magnitude across about one-third of the foil surface area as shown in Figure 11 (4b)-(4d).

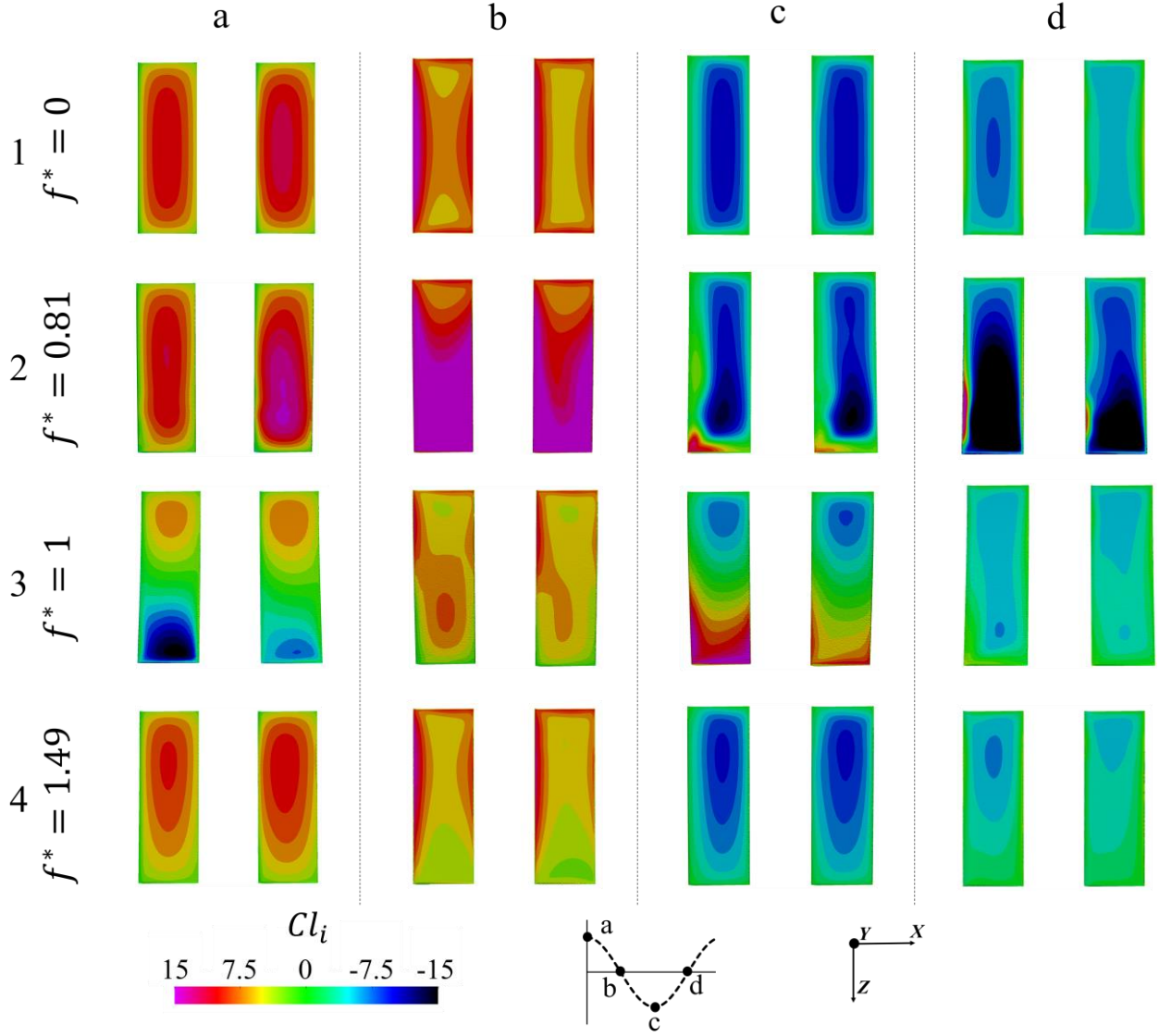

Figure 12 Surface contour of the facial lift coefficient at St = 0.45 with (a) $t/T = 0$, (b) $t/T = 0.25$, (c) $t/T = 0.5$, (d) $t/T = 0.75$, (1) $f^* = 0$, (2) $f^* = 0.81$, (3) $f^* = 1$ and (4) $f^* = 1.49$.

As in Eq. (12), the power consumed by the heaving foil is calculated by the production of the foils' lift and the foils' heaving velocity. The facial lift coefficient of cell face $i$ in the fluid domain is defined as

$$Cl_i = \frac{lift_i}{\frac{1}{2}\rho_f u^2 A_i}, \quad (27)$$

where, $lift_i$ is the lift force applied on the $i$th cell face. Figure 12 shows the surface contour of the facial lift coefficient at $St = 0.45$ with four reduced frequencies at four time instances. A high lift region at the free tip with a low lift region next to it for both fore foil and hind foil is observed in Figure 12 (2c). It is observed at the Figure 12 (2a) that the lift contour of the fore foil with $f^* = 0.81$ is close to that of the fore foil with $f^* = 0$ as in Figure 12 (1a). A high lift force region is observed from the hind foil with $f^* = 0.81$. It does not show from the hind foil with $f^* = 0$. At $t/T = 0.75$, a high-pressure region is generated around the free tip of the upper and low surface, respectively, for both the fore foil and the hind foil with $f^* = 0.81$.

Figure 12 (2b) and (2d) shows that the lift around the free tip for both the fore foil and the hind foil is much larger than that of the foils in Figure 12 (1b) and (1d).

At $t/T = 0.5$, the forced heaving tip is at the beginning of its up-stroke, but the free tip is still under its down-stroke for both the fore foil and the hind foil with $f^* = 1$ (due to the $130°$ phase difference between tips; see Figure 7). As a result, the half foil close to the forced heaving tip has a different sign of $Cl_i$ compared with that of the other half of the foil for both of the dual foils at $t/T = 0$ (Figure 12 (3a)) and $t/T = 0.5$ (Figure 12 (3c)). At $t/T = 0.75$, both the forced heaving tip and the free tip are on their up-stroke. The forced heaving tip is in the middle of its up-stroke, but the free tip is just at the beginning of its up-stroke. Figure 12 (3b) and (3d) show that most areas of the foil surface have the same sign of $Cl_i$ for both the fore foil and the hind foil. There is a positive $Cl_i$ area (see Figure 12 (3d)) can be seen at the free tip of the fore foil. As shown in Figure 7 (2), the heaving amplitude of the free tip for the hind foil is smaller than that of the fore foil at $f^* = 1$, which leads to the $|Cl_i|$ at the half foil close to the free tip of the hind foil being smaller than that of the fore foil for the during the heaving circle as shown in Figure 12 (3).

Figure 7 (3) demonstrates that the heaving motion of the free tip has a smaller amplitude with no phase difference compared with that of the forced heaving tip with $f^* = 1$. Figure 12 (4) shows that the facial lift coefficient around the forced heaving tip of the fore foil and the hind foil at $f^* = 1.49$ are close with that of the fore foil and hind foil at $f^* = 0$. As a result of the smaller sectional amplitude along the spanwise of the foils, the absolute values of the facial lift coefficient around the free tip of the fore foil and hind foil at $f^* = 1.49$ are smaller than that of the fore foil and hind foil at $f^* = 0$.

It can be concluded that the difference in the propulsion efficiency between the flexible dual foils compared with that of the single rigid foil under the present working condition is impacted by three main factors, which are the vortices due to the foil interaction, the foil maximum effective angle of attack and the phase difference between the foil forced heaving tip and the free tip. Both thrust and lift generation are affected by the heaving amplitude (i.e. the $\overline{\langle \alpha_{eff} \rangle_{max}}$) and the phase difference between the forced heaving and free tips of the foil. The thrust (or drag) generation is also affected by the interaction between the foils.

### 3.2.5. Elastic Behaviour at $St = 0.64$ and $St = 0.95$

In this subsection, we will examine the elastic behaviour of dual heaving foils in two key regimes: at peak propulsion efficiency ($St = 0.64$) and in the post-peak propulsion efficiency regime ($St = 0.95$). At the peak propulsion efficiency point at $St = 0.64$, rigid foils have already achieved an optimal effective angle of attack (i.e. $\langle \alpha_{eff} \rangle_{ref}$). Introducing flexible foils with elastic displacement causes $\alpha_{eff}$ to deviate from the optimal range, resulting in lower propulsion efficiency compared to rigid foils. In the post-peak propulsion efficiency regime such as $St = 0.95$, where $\alpha_{eff}$ of rigid foils surpasses the optimum effective angle of attack, enhanced propulsion efficiency can be achieved by reducing heaving amplitude while maintaining a small phase difference between the forced heaving tip and the free tip when employing flexible foils.

Figure 13 shows the normalised mid-chord position of flexible dual heaving foils for three reduced frequencies at $St = 0.64$ and $St = 0.95$. The normalised mid-chord position of flexible dual heaving foils at $St = 0.64$ and $St = 0.95$ have similar behaviour in terms of the sectional heaving amplitude along the spanwise direction and the phase difference between tips as that of $St = 0.45$. More precisely, the sectional amplitude of the heaving motion gradually increases from the forced heaving tip to the free tip at $f^* = 0.81$. When the reduced frequency of the foils is 1, the sectional amplitude of the heaving motion decreases from the forced heaving tip until the mid-span, and then increases towards the free tip. When $f^* = 1.49$, the sectional amplitude of the heaving motion gradually decreases from the forced heaving tip to the free tip. Phase lags between foil tips can be observed at $f^* = 0.81$ and $f^* = 1$, and no noticeable phase lag can be observed between tips of the foil at $f^* = 1.49$.

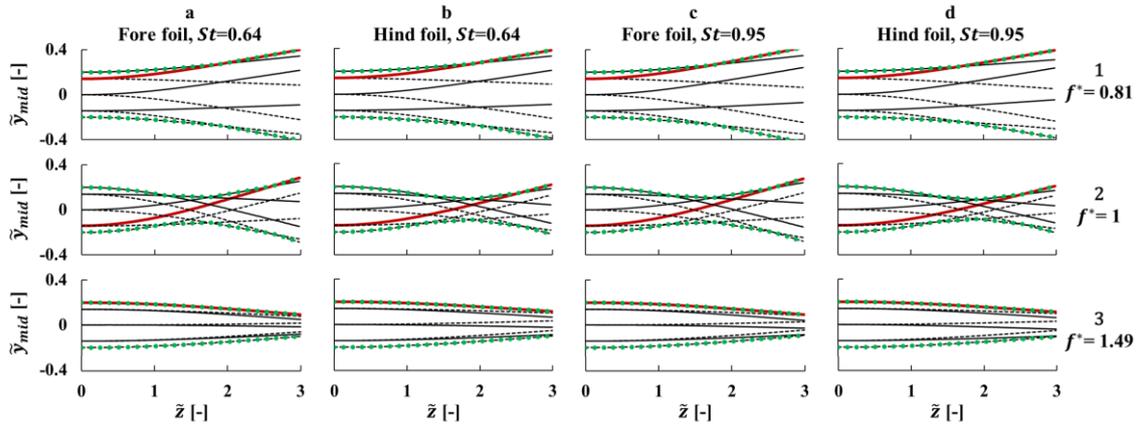

Figure 13 Mid-chord position of flexible dual heaving foils that is normalised by chord at three reduced frequencies with (a) fore foil at $St = 0.64$, (b) hind foil at $St = 0.64$, (c) fore foil at $St = 0.95$, (d) hind foil at $St = 0.95$, (1) $f^* = 0.81$, (2) $f^* = 1$ and (3) $f^* = 1.49$. The $x$-axis represents the normalised position along $z$-axis with a $\frac{3c}{2}$ offset towards the positive direction. The solid black line represents the down-stroke motions. The dashed black line represents the up-stroke motions. The green dotted line represents the envelope of foil deformation. The red solid line represents the mid-chord position with the maximum deformation.

Figure 14 shows the normalised maximum effective angle of attack and phase difference between the forced heaving tip and the free tip at four reduced frequencies at $St = 0.64$ and $St = 0.95$. The normalised maximum effective angle of attack of both the fore foil and the hind foil at $f^* = 0.81$ are higher than 1, while that at $f^* = 1$ and $f^* = 1.49$ they are all smaller than 1 at $St = 0.64$. The normalised maximum effective angle of attack of the flexible dual foils ($f^* \neq 0$) at $St = 0.64$ are all farther away from 1, which is the reason of the dual foils at $St = 0.64$ not having an efficiency gain by using flexibility foils compared with that of the rigid foils, which is different from that at $St = 0.45$.

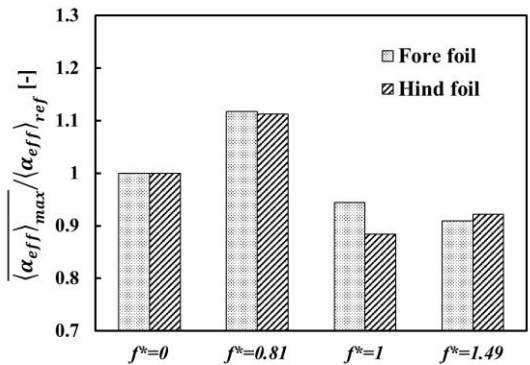

(a) Normalised maximum effective angle of attack at $St = 0.64$

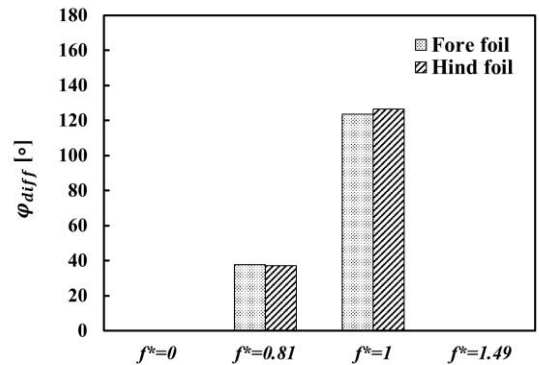

(b) The phase difference between the forced heaving tip and the free tip at $St = 0.64$

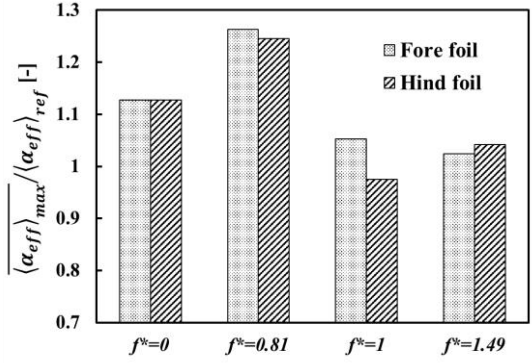
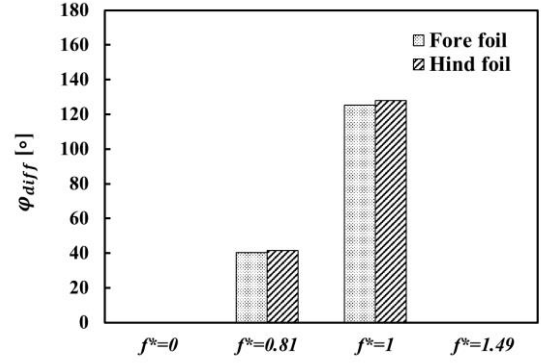

(c) Normalised maximum effective angle of attack at $St = 0.95$

(d) The phase difference between the forced heaving tip and the free tip at $St = 0.95$

Figure 14 Normalised maximum effective angle of attack (a and c) and phase difference between the forced heaving tip and the free tip (b and d) at four reduced frequencies at $St = 0.64$ (a and b) and $St = 0.95$ (c and d).

At $St = 0.95$, as shown in Figure 14 (c), the normalised maximum effective angle of attack is further away from 1 at $f^* = 0.81$ compared with that of $f^* = 0$, which leads to an efficiency loss as can be seen from Figure 4 (a) and Figure 5 (c). When $f^* = 1$ and $f^* = 1.49$, the normalised maximum effective angle of attacks is closer to 1 than that of $f^* = 0$. As we have observed in Figure 4 (a) and Figure 5 (c), the propulsion efficiency of the dual foils at $St = 0.95$ and $f^* = 1$ is lower than that of the dual foils at $St = 0.95$ and $f^* = 0$. It is due to the impact of the phase difference (about 130°) between the forced heaving tip and the free tip for both the fore foil and the hind foil at $f^* = 1$, as can be observed in Figure 14 (d). Without the impact of the phase difference between foil tips, there is a gain on the propulsion efficiency of the dual foils at $St = 0.95$ and $f^* = 1.49$, as can be seen from Figure 4 (a) and Figure 5 (c).

### 3.3. Flexible Dual Foils with Fixed Reduced Frequency and Fixed Natural Frequency

Here we consider the propulsion performance of flexible dual heaving foils at different Strouhal numbers under two reduced frequencies and three natural frequencies.

To study the effect of the Strouhal number on flexible dual heaving foils at a fixed reduced frequency, the natural frequency of the foils, along with the heaving frequency (or Strouhal numbers), is changed. This is a straightforward task in numerical simulation but is challenging to achieve experimentally. Two reduced frequencies, $f^* = 0.81$ and $f^* = 1$, are selected since the dual heaving foils will achieve the best and the worst propulsion performance under these reduced frequencies, respectively, according to the results in Section 3.2.

Figure 15 shows the propulsion efficiency increment, thrust coefficient increment, and power coefficient increment, of flexible dual heaving foils compared with that of rigid foils with Strouhal number ranging from 0.4 to 1 at $f^* = 0.81$ and $f^* = 1$. When $f^* = 0.81$, a propulsion efficiency gain is observed using the flexible material compared with that of rigid foils at $St < 0.6$. The smaller the Strouhal number, the greater the propulsion efficiency gain. As discussed in Section 3.2, the gain is due to the impact of the flexible motion of the heaving foils, which makes $\frac{\langle \alpha_{eff} \rangle_{max}}{\langle \alpha_{eff} \rangle_{ref}}$ approach 1 at small Strouhal numbers. When $St > 0.6$, the propulsion efficiency increments of dual heaving foils using flexible material becomes negative. It indicates a propulsion efficiency loss compared with that of rigid heaving foils as the flexible motion of the heaving foils pushes $\frac{\overline{\langle \alpha_{eff} \rangle_{max}}}{\langle \alpha_{eff} \rangle_{ref}}$ further away from 1. As shown in Figure 15 (b) and (c), the larger the Strouhal number, a larger thrust coefficient increment and power coefficient increment can be achieved when a flexible material is used for dual heaving foils with $f^* = 0.81$, compared with that of rigid dual heaving foils. It is also apparent that the growth rate of the power coefficient increment is greater than the growth rate of the thrust coefficient increment when $St > 0.6$ at $f^* = 0.81$. When $f^* = 1$, negative propulsion efficiency increments are observed over the entire study interval of the Strouhal number. As discussed in Section 3.2, it is the combined effect of the normalised maximum effective angle of attack as well as the phase difference between the forced heaving tip and the free tip. As in Figure 15 (b) and (c), the larger the Strouhal number, the smaller the thrust coefficient increment and power coefficient increment for flexible dual heaving foils with $f^* = 1$, compared with that of rigid dual heaving foils.

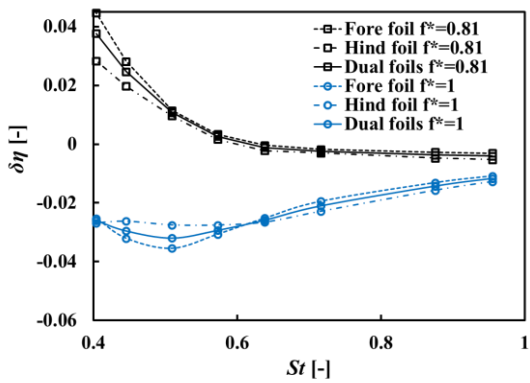

(a) Propulsion efficiency increment compared with that of rigid foils

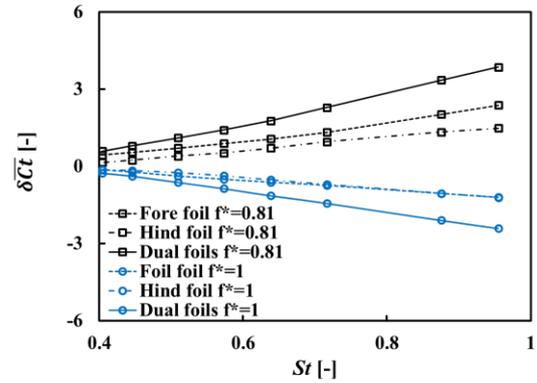

(b) Thrust coefficient increment compared with that of rigid foils

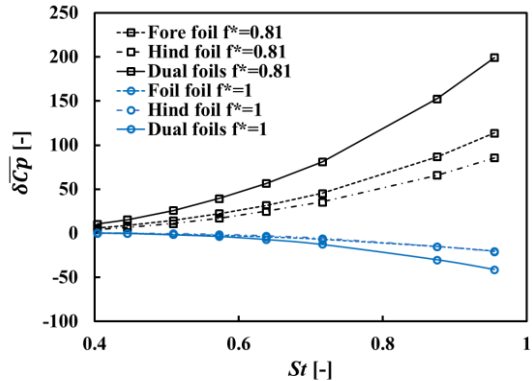

(c) Power coefficient increment compared with that of rigid foils

Figure 15. Propulsion performance of flexible dual heaving foils for $0.4 < St < 1$ at two reduced frequencies with (a) propulsion efficiency increment, (b) thrust coefficient increment and (c) power coefficient increment compared with that of rigid foils.

In practice, this is what will happen in reality and can be repeated by the experimental analysis when the flexible dual heaving foils are subject to a fixed natural frequency but different Strouhal numbers. Under this condition, the Young's modulus of the heaving foil remains the same while the Strouhal number changes. The resonance Strouhal number, $St_r$, is defined as the Strouhal number with respect to the heaving motion of the dual heaving foils when resonance occurs (i.e. the reduced frequency approaching 1). Three resonance Strouhal numbers are studied ($St_r = 0.64$, $St_r = 0.78$ and $St_r = 1.17$).

Figure 16 shows the propulsion efficiency increment and thrust coefficient increment of flexible dual heaving foils with Strouhal numbers ranging from 0.4 to 1 at three resonance Strouhal numbers. When the resonance Strouhal number equals 0.64, there is a positive propulsion efficiency increment and thrust coefficient increment at $St < 0.5$ (i.e. $f^* < 0.8$) compared with that of rigid dual heaving foils. When St approaches $St_r$, both propulsion efficiency increment and thrust coefficient increment drop to negative values. Further increasing $St$ lead to a gradually increasing of the propulsion efficiency increment, while the thrust coefficient increment remains stable, as can be seen from Figure 16 (a) and (b). The propulsion efficiency increment and thrust coefficient increment follow the same trend when $St_r = 0.78$ and $St_r = 1.17$ as in Figure 16 (c)-(f). It can be concluded from Figure 16 that the dual heaving foils with a larger resonance Strouhal number can achieve a greater thrust increment compared with that of rigid dual heaving foils when working at the reduced frequency of about 0.8. While the dual heaving foils with a smaller resonance Strouhal number can achieve a larger propulsion efficiency increment compared with that of rigid dual heaving foils when working at small Strouhal numbers.

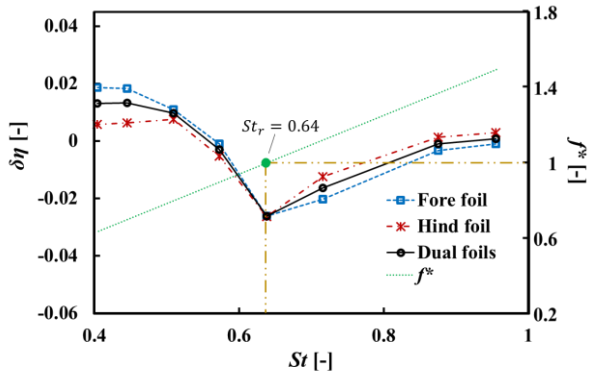

(a) Propulsion efficiency increment compared with that of rigid foils at $St_r = 0.64$

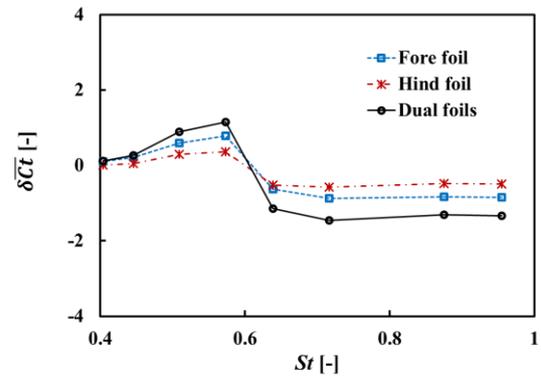

(b) Thrust coefficient increment compared with that of rigid foils at $St_r = 0.64$

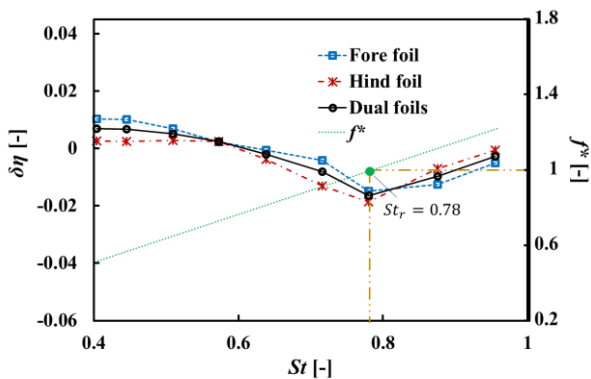

(c) Propulsion efficiency increment compared with that of rigid foils at $St_r = 0.78$

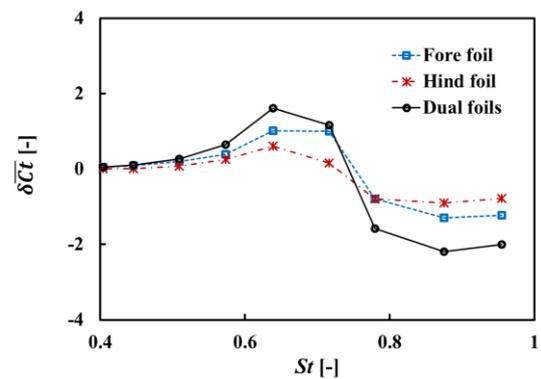

(d) Thrust coefficient increment compared with that of rigid foils at $St_r = 0.78$

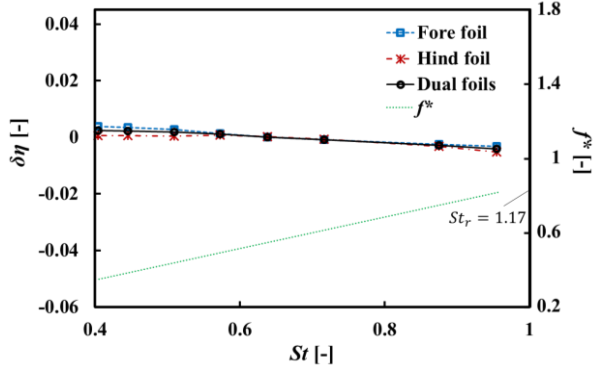
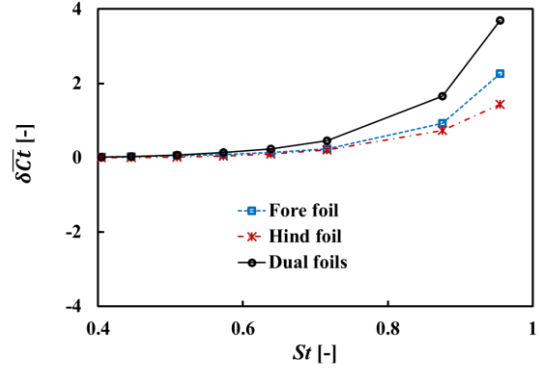

(e) Propulsion efficiency increment compared with that of rigid foils at $St_r = 1.17$

(f) Thrust coefficient increment compared with that of rigid foils at $St_r = 1.17$

Figure 16. Propulsion performance of flexible dual heaving foils for $0.4 < St < 1$ at (a) and (b) $St_r = 0.64$, (c) and (d) $St_r = 0.78$ and (e) and (f) $St_r = 1.17$ with (a), (c) and (e) propulsion efficiency increment and (b), (d) and (f) thrust coefficient increment compared with that of rigid foils.

## 4. Conclusions

The impact of spanwise flexibility on the propulsion performance of three-dimensional flexible dual foils with prescribed sinusoidal heaving motion is studied with the Reynolds number set to 100, based on published literature. A decoupled analysis has been carried out, by fixing the density of the foil structure while varying its natural frequency. A wide range of natural frequencies covering pre-resonance, resonance and post-resonance regimes are studied. We have also investigated the propulsion performance of the flexible dual heaving foils under different Strouhal numbers with either fixed reduced frequency or fixed natural frequency, as in real working conditions.

We observe significant differences between the propulsion efficiency of the rigid fore foil, rigid hind foil and the rigid single foil when a Strouhal number smaller than 0.64 is used. Conversely, negligible differences between the rigid fore and hind foils and the single foil are observed at larger Strouhal numbers. By comparing the propulsion efficiency of dual heaving foils with a spacing of 1 chord with that of dual heaving foils with infinite spacing, we observe a loss in propulsion efficiency for the former at the small Strouhal number, which is due to the thrust loss of the hind foil caused by the interaction between the fore foil and the hind foil. Through analysis of the flow structure of the dual heaving foils, we conclude that when the leading edge vortex of the hind foil and trailing edge vortex of the fore foil have the same sign (in terms of their spanwise vorticity), there will be a thrust enhancement at the leading edge of the hind foil, and vice-versa. We also observe trailing edge vortex breakup due to the presence of the hind foil.

We subsequently investigate the propulsion performance of *flexible* dual heaving foils with different reduced frequencies at three Strouhal numbers. The propulsion efficiency, thrust force and power consumption drop (relative to the rigid foils) are analysed, while the tip deformation coefficient increases close to the resonance regime. When the Strouhal number is 0.45, the propulsion efficiency of the flexible dual foils has a significant increase when the reduced frequency increases from 0 to 0.81. Both cycle averaged thrust and the cycle averaged power coefficient of flexible dual heaving foils increase with the increase of the reduced frequency during the pre- and post-resonance regimes. An instability is observed in the instantaneous tip deformation coefficient of the flexible dual foils at the post-resonance regimes. Different sectional amplitudes of the heaving foil along the spanwise direction as well as the phase lag between tips of the foil can be observed for different reduced frequency regimes. The impact of flexibility can be predicted by the normalised maximum effective angle of attack as well as the phase difference between the foil tips. We conclude that the impact of flexibility dominates the fore foil at all three reduced frequency regimes, while both the flexibility and the interactions between foils have a great impact on the flexible hind foil. Both thrust generation and lift generation are affected by the maximum value of the effective angle of attack of the foil over a heaving cycle and the phase difference between the forced heaving and free tips of the foil. The thrust (or drag) generation is also affected by the interaction between the foils.

We then investigate the propulsion performance of flexible dual heaving foils at different Strouhal numbers with fixed reduced frequency, $f^*$. When $f^* = 0.81$, a gain in propulsion efficiency in the flexible configuration is observed, relative to rigid foils, for Strouhal numbers less than 0.6. Both the thrust coefficient increment and the power coefficient increment are proportional to the Strouhal number when a flexible material is used for dual heaving foils with $f^* = 0.81$ compared with that of rigid dual heaving foils. Negative propulsion efficiency increments, thrust coefficient increment and the power coefficient increment are present over the entire study interval of the Strouhal number at the resonance regime. We can conclude from the analysis of the flexible dual heaving foils under a fixed natural frequency but different Strouhal numbers that the dual heaving foils with a larger resonance Strouhal number can achieve a greater thrust increment compared with that of rigid dual heaving foils when working at the reduced frequency about 0.8. While the dual heaving foils with a smaller resonance Strouhal number can achieve a larger propulsion efficiency increment compared with that of rigid dual heaving foils when working at small Strouhal numbers. Although optimising the dual heaving foils for a better propulsion performance is not a scope of this study, there is no doubt that the findings and conclusions of this study will contribute and provide guidance on these optimising activities in the future.

Future works of this study include extending the working conditions to high Reynolds number regimes and considering high efficiency self-propelled flexible propulsor. The propulsion performance of dual heaving foils with other foil shapes will also be investigated in our future studies.

**Supplementary: Flow Field Analysis of Rigid Foil Propulsion**

To investigate the mechanism of the interaction between the rigid dual foils in terms of the propulsion performance, we study three Strouhal numbers ($St = 0.45$, 0.64 and 0.95, respectively). The surface contour of facial thrust coefficient and iso-surface of $Q$-criterion of a single heaving foil with $St = 0.45$ is shown in Figure 17. The Leading Edge Vortices (LEVs) are related to the thrust generation, while the Trailing Edge Vortices (TEVs) are related to drag generation. There is a clear separation between positive and negative thrust coefficients at the surface of the airfoil, as illustrated in Figure 17 (a)-(d). The total thrust of the foil is found by integrating the thrust across the upper and lower surface of the foil. Note that, due to the symmetry of the problem, the distribution of the thrust coefficient for the upper surface has a 180° phase shift with respect to the heaving cycle, compared with that of the lower surface, i.e. the facial thrust coefficient distribution of the lower surface of Figure 17 (a) is identical with that of Figure 17 (c) and the same for Figure 17 (b) and Figure 17 (d).

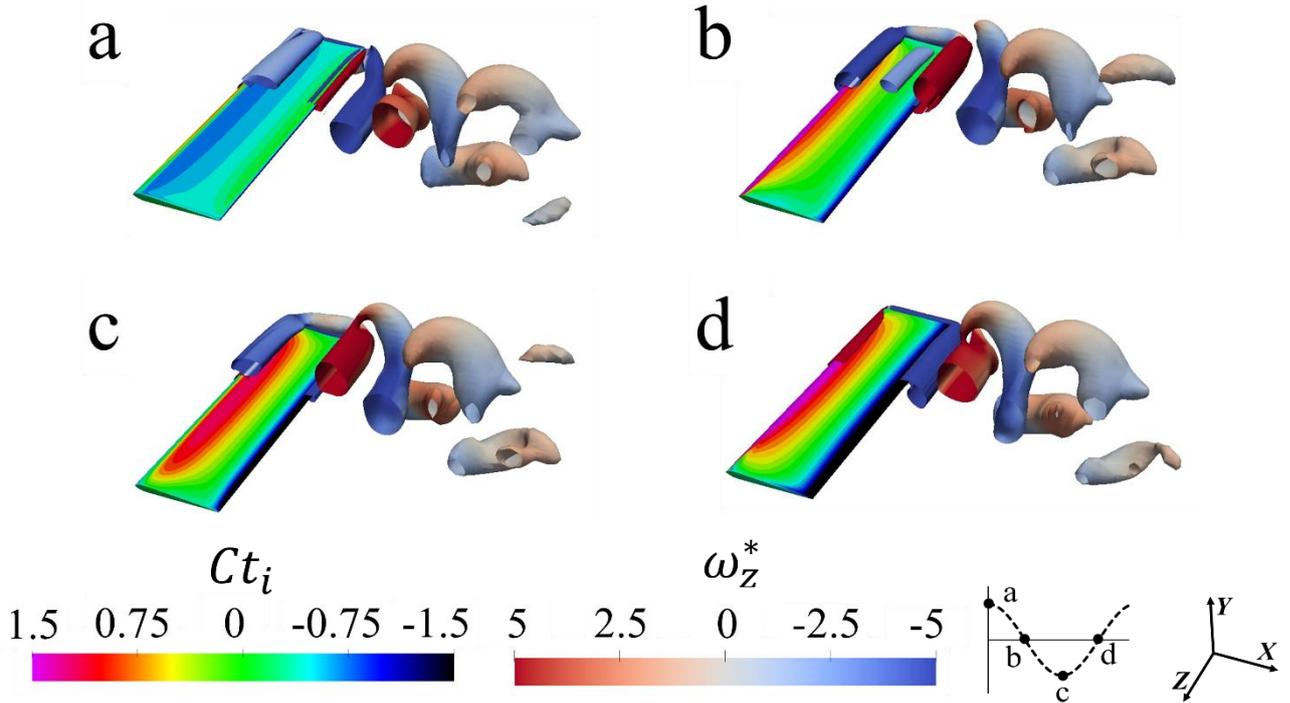

Figure 17 Surface contour of the facial thrust coefficient and iso-surface of $Q$-criterion ($\frac{Qc^2}{u^2} = 1$) coloured by the normalised $z$-vorticity on single heaving foil at $St = 0.45$ with (a) $t/T = 0$, (b) $t/T = 0.25$, (c) $t/T = 0.5$ and (d) $t/T = 0.75$.

We further plot the surface contour of the facial thrust coefficient and iso-surface of $Q$-criterion of the dual heaving foils with $St = 0.45$ as in Figure 18. Compared with that of the single heaving foil in Figure 17, we can observe the impact of the TEV of the fore foil on the formation of the LEV and the thrust generation of the hind foil. With the presence of the fore foil's TEV with a negative sign of its vorticity, the vorticity of the hind foil's LEV (with a negative sign) is stronger and the facial thrust coefficient at the leading edge of the upper surface of the hind foil is larger than that of the fore foil at $t/T = 0$ as shown in Figure 18 (a), and is also larger than that of the single foil at $t/T = 0$ as shown in Figure 17 (a). We also observe drag reduction around one third of the span of the upper surface of the hind foil compared with that of the fore foil. While at $t/T = 0.25$ as shown in Figure 18 (b), the hind foil's LEV with negative sign has a flatter cross-sectional shape and the vorticity is weaker than that of the fore foil with the existence of the fore foil's TEV with a positive sign. It leads to a thrust reduction around the mid-span of the leading edge of the hind foil's upper surface compared with that of the fore foil and the single foil, as shown in Figure 18 (b). A similar phenomenon can be observed at $t/T = 0.5$ and $t/T = 0.75$ in terms of the interaction between the fore foil's TEV and the hind foil's LEV as well as the thrust reduction on the leading edge of the hind foil, as can be seen from Figure 18 (c)-(d). We can conclude that if the LEV of the hind foil and TEV of the fore foil have the same sign in terms of their vorticity (as Figure 18 (a) shows), there will be a thrust enhancement at the leading edge of the hind foil. Otherwise, if the LEV of the hind foil and TEV of the fore foil have the opposite sign in terms of their vorticity, there will be a thrust reduction at the hind foil's leading edge (as Figure 18 (b)-(d) shows). We also observe that the TEV of the fore foil is split into two parts due to the existence of the hind foil as indicated in Figure 18 (a).

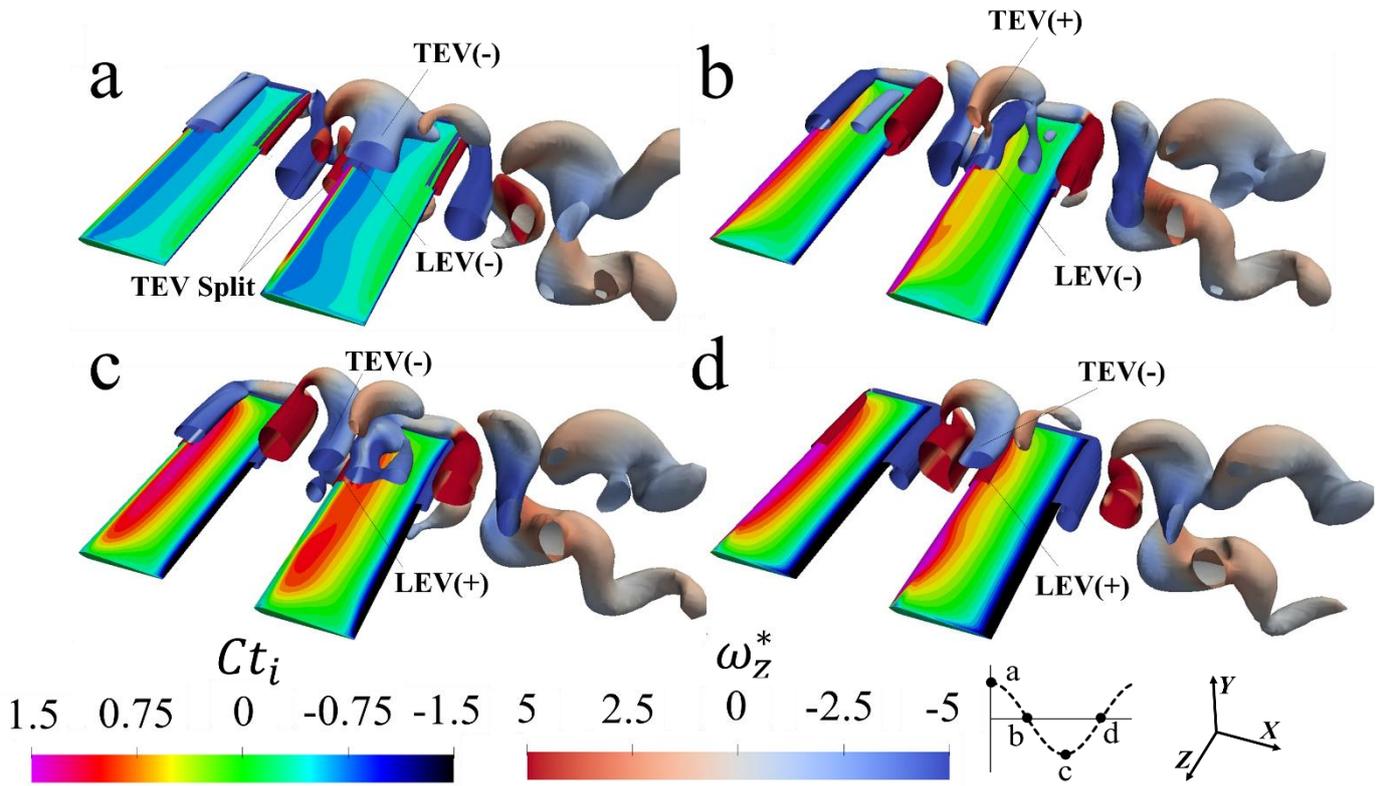

Figure 18 Surface contour of the facial thrust coefficient and iso-surface of $Q$-criterion ($\frac{Qc^2}{u^2} = 1$) coloured by the normalised $z$-vorticity on dual heaving foils at $St = 0.45$ with (a) $t/T = 0$, (b) $t/T = 0.25$, (c) $t/T = 0.5$ and (d) $t/T = 0.75$.

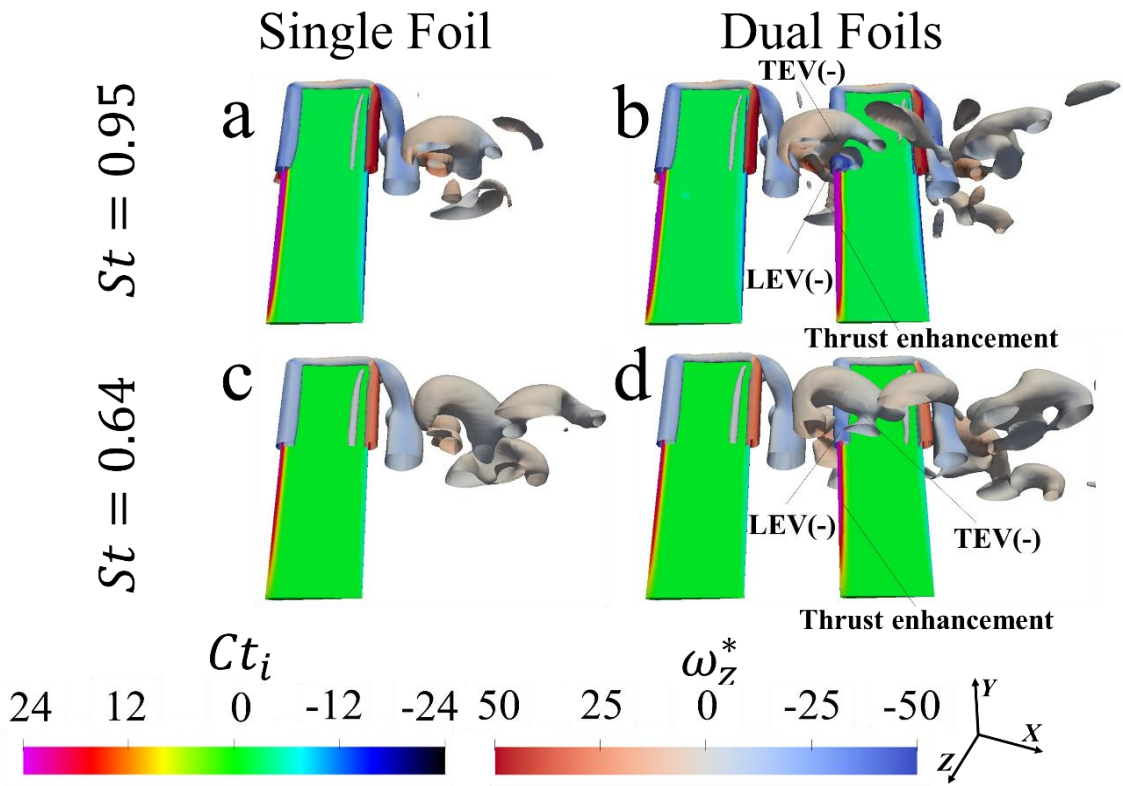

Figure 19 Surface contour of the facial thrust coefficient and iso-surface of $Q$-criterion ($\frac{Qc^2}{u^2} = 100$) coloured by the normalised $z$-vorticity at $t/T = 0$ with (a) and (b) $St = 0.95$, (c) and (d) $St = 0.64$, (a) and (c) single heaving foil and (b) and (d) dual heaving foils.

The phenomenon of the thrust enhancement and thrust reduction at the leading edge of the hind foil also occurs when the Strouhal number of heaving foils is larger than 0.45. Figure 19 and Figure 20 show the surface contour of the facial thrust coefficient and iso-surface of $Q$-criterion with both $St = 0.64$ and $St = 0.95$ for the single foil and the dual foils at $t/T = 0$ and $t/T = 0.75$, respectively. In Figure 19, the thrust is generated around the mid-span of the leading edge of the hind foil is larger than that of the fore foil as well as that of the single foil due to the interaction between the hind foil's LEV with negative vorticity and the fore foil's TEV with negative vorticity. While in Figure 20, no noticeable difference can be observed between the LEV of the fore foil and the LEV of the single foil in terms of their shape and strength for both $St = 0.64$ and $St = 0.95$ at $t/T = 0.75$. The vorticity of LEV with positive vorticity of the hind foil for both $St = 0.64$ and $St = 0.95$ at $t/T = 0.75$ are smaller compared with that of the single foil due to the presents of the fore foil's TEV with negative vorticity. It leads to a thrust reduction at the upper surface of the hind foil as indicated in Figure 20 (b) and (d).

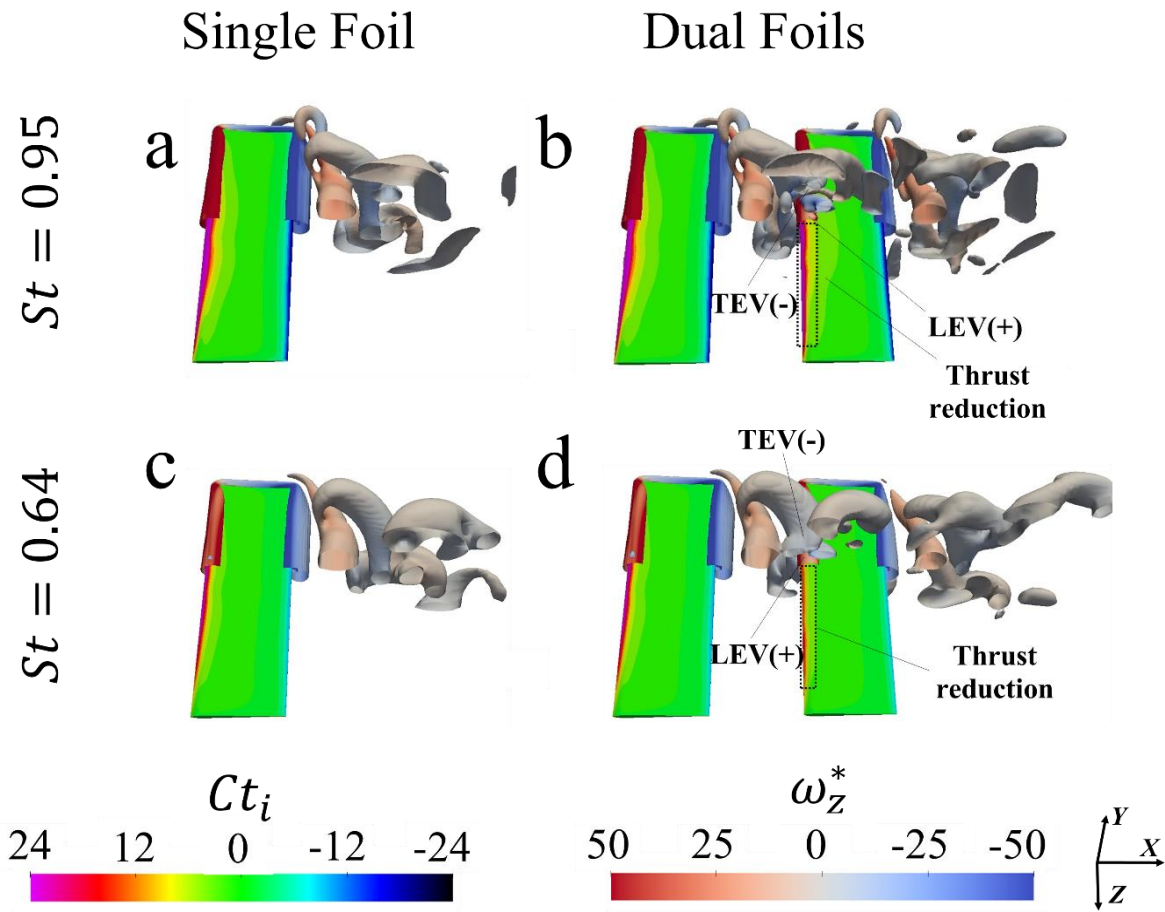

Figure 20 Surface contour of the facial thrust coefficient and iso-surface of $Q$-criterion ($\frac{Qc^2}{u^2} = 100$) coloured by the normalised $z$-vorticity ($\omega_z^*$) at $t/T = 0.75$ with (a) and (b) $St = 0.95$, (c) and (d) $St = 0.64$, (a) and (c) single heaving foil and (b) and (d) dual heaving foils.

The surface contours of facial lift coefficient at $St = 0.45$ for both the single foil and the dual foils are shown in Figure 21. The $Cl_i$ for the upper surfaces of the fore foil and the hind foil at

$t/T = 0$ and $t/T = 0.5$ are quite close to that of the single foil. While at $t/T = 0.25$ and $t/T = 0.75$, the absolute value of $Cl_i$ of the mid-span area of the upper surface of the fore foil is larger than that of the single foil. The absolute value of $Cl_i$ of the mid-span area of the upper surface of the hind foil is smaller than that of the single foil. It is consistent with what we observe from Figure 3 (b) in the main article that the instantaneous power coefficients for the fore foil, hind foil and the single foil have little difference at $t/T = 0$ and $t/T = 0.5$, but have significant differences at $t/T = 0.25$ and $t/T = 0.75$.

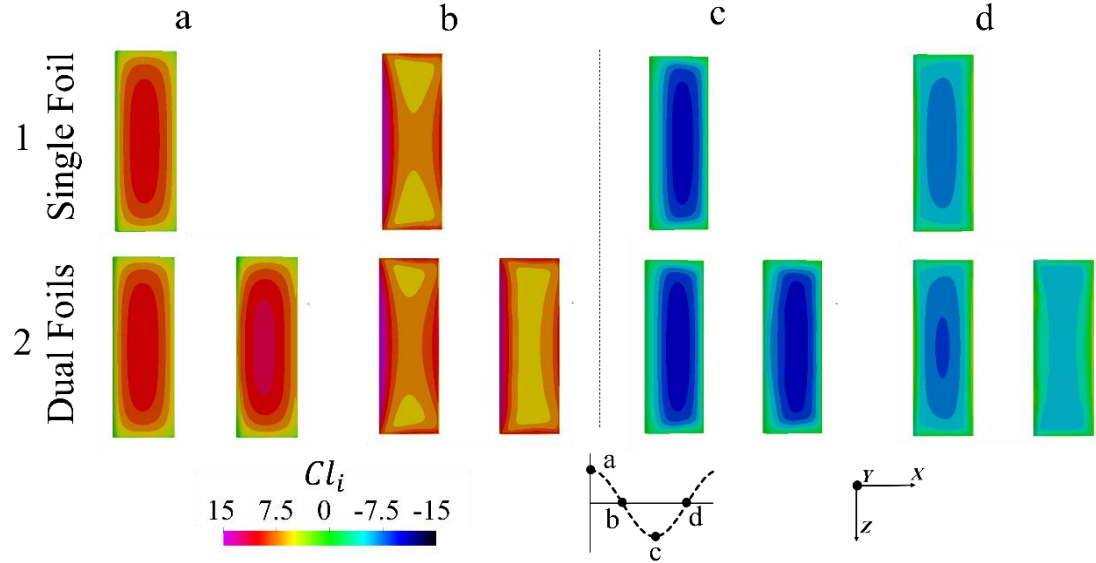

Figure 21 Surface contour of the facial lift coefficient at $St = 0.45$ with (a) $t/T = 0$, (b) $t/T = 0.25$, (c) $t/T = 0.5$, (d) $t/T = 0.75$, (1) single heaving foil and (2) dual heaving foils.

The pressure coefficient is defined as

$$C_{pre} = \frac{p}{\frac{1}{2}\rho_f u^2}. \tag{28}$$

We plot the surface contour of pressure coefficient for both the single foil and the dual foils with $St = 0.45$ at $t/T = 0.25$ and $t/T = 0.75$ in Figure 22. The fluid pressure distribution around the mid-span is quite different among these three foils at $t/T = 0.25$ and $t/T = 0.75$, leading to the difference in the lift generation as well as the power coefficient at these time instances as in Figure 21 and Figure 3 (b) in the main article.

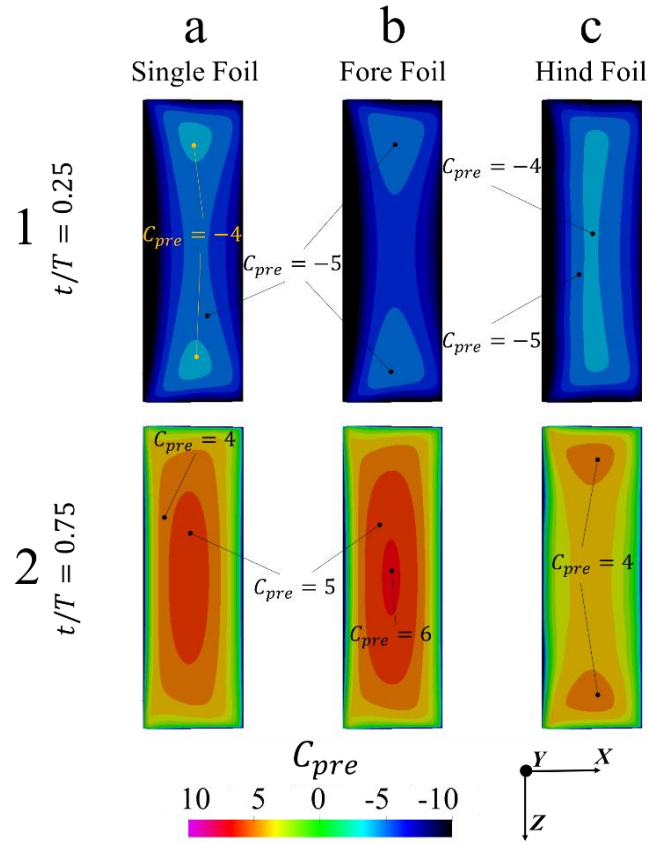

Figure 22 Surface contour of the pressure coefficient at $St = 0.45$ with (a) single foil, (b) fore foil of the dual foils, (c) hind foil of the dual foils, (1) $t/T = 0.25$ and (2) $t/T = 0.75$.

These results show that the interactions between the dual foils have a greater impact on the hind foil than its impact on the fore foil in terms of their propulsion performance and flow structure.

**Acknowledgements**

This work was supported by the STFC Hartree Centre's Innovation Return on Research programme, funded by the Department for Business, Energy & Industrial Strategy. Results were obtained using the Scafell Pike High-Performance Computer based at the STFC Hartree Centre. The coupling between OpenFOAM and MUI library used in this study was derived from the original implementation by Dr Stephen M. Longshaw. The coupling between the FEniCS library and the MUI Python wrapper was derived from the original implementation by Dr Eduardo R. Fernandez.